\documentclass[aps,manuscript,showpacs,showkeys,superscriptaddress]{revtex4-1}
\usepackage{graphicx}
\usepackage{epsfig}  
\usepackage{epsf}    
\usepackage{dcolumn}
\usepackage{bm}
\usepackage{dcolumn}
\usepackage{textcomp}
\usepackage[tbtags]{amsmath}
\usepackage{amsfonts}
\usepackage{float}
\usepackage{subfig}
\usepackage{color}
\usepackage[]{hyperref}
  \hypersetup{
  unicode=false,          
  pdftoolbar=true,        
  pdfmenubar=true,        
  pdffitwindow=true,     
  pdfstartview={FitH},    
  pdfsubject={Getting the best out of T2K and NOvA},   
  pdfnewwindow=true,      
  pdfcreator={RevTeX},
  colorlinks=true,       
  linkcolor=red,          
  citecolor=blue,        
  urlcolor=blue,           
  }
\usepackage{hypcap}

\def\anu{{\bar\nu}}

\newcommand{\beq}{\begin{equation}}
\newcommand{\eeq}{\end{equation}}
\newcommand{\beqa}{\begin{eqnarray}}
\newcommand{\eeqa}{\end{eqnarray}}

\newcommand{\tx}{{\theta_{12}}}
\newcommand{\ty}{{\theta_{13}}}
\newcommand{\tz}{{\theta_{23}}}

\newcommand{\dl}{{\Delta_{31}}}
\newcommand{\ds}{{\Delta_{21}}}

\newcommand{\ahat}{\hat{A}}
\newcommand{\dhat}{\hat{\Delta}}

\newcommand{\dcp}{\delta_{\mathrm{CP}}}
\newcommand{\nova}{NO$\nu$A~}

\newcommand{\pme}{P_{\mu e}}

\newcommand{\pmebar}{P_{\bar{\mu} \bar{e}}}

\newcommand{\dchsq}{\Delta\chi^2}

\newcommand{\red}{\textcolor{red}}

\begin{document}
\title{Searching for non-unitary neutrino oscillations in the present T2K and NO$\nu$A data}

\author{Luis Salvador Miranda}
\affiliation{Centre for Astro-Particle Physics (CAPP) and Department of Physics, University of Johannesburg, PO Box 524, Auckland Park 2006, South Africa}
\author{Pedro Pasquini}
\affiliation{Instituto  de F\'isica Te\'orica--Universidade Estadual Paulista (UNESP)\\
    R. Dr. Bento Teobaldo Ferraz 271, Barra Funda, S\~ao Paulo - SP, 01140-070, Brazil}
    
    \affiliation{Tsung-Dao Lee Institute \& School of Physics and Astronomy, Shanghai Jiao Tong University, China}
\author{Ushak Rahaman}
\affiliation{Centre for Astro-Particle Physics (CAPP) and Department of Physics, University of Johannesburg, PO Box 524, Auckland Park 2006, South Africa}
\author{Soebur Razzaque}
\affiliation{Centre for Astro-Particle Physics (CAPP) and Department of Physics, University of Johannesburg, PO Box 524, Auckland Park 2006, South Africa}

%
\date{Received: date / Revised version: date}
%
\begin{abstract}
The mixing of three active neutrino flavors is parameterized by the unitary PMNS matrix. If there are more than three neutrino flavors and if the extra generations are heavy iso-singlets, the effective $3\times 3$ mixing matrix for the three active neutrinos will be non-unitary. We have analyzed the latest T2K and \nova data with the hypothesis of non-unitary mixing of the active neutrinos. We found that the 2019 \nova data slightly (at $\sim 1\, \sigma$ C.L.) prefer the non-unitary mixing over unitary mixing. In fact, allowing the non-unitary mixing brings the \nova best-fit point in the $\sin^2\tz-\dcp$ plane closer to the T2K best-fit point. The 2019 T2K data, on the other hand, cannot rule out any of the two mixing schemes. A combined analysis of the \nova and T2K 2019 data prefers the non-unitary mixing at $1\, \sigma$ C.L.. We derive constraints on the non-unitary mixing parameters using the best-fit to the combined \nova and T2K data. These constraints are weaker than previously found. The latest 2020 data from both the experiments prefer non-unitarity over unitary mixing at $1\, \sigma$ C.L. The combined analysis prefers non-unitarity at $2\, \sigma$ C.L. The stronger tension, which exists between the latest 2020 data of the two experiments, also gets reduced with non-unitary analysis.
\end{abstract} 
\maketitle
\section{Introduction}
Neutrino oscillation phenomena has been well established by the solar \cite{Bahcall:2004ut, Ahmad:2002jz}, atmospheric \cite{Fukuda:1994mc} and reactor \cite{An:2012eh, Ahn:2012nd, Abe:2011fz} neutrino experiments. The oscillations between 3-neutrino flavors depend on three mixing angles $\tx$, $\ty$ and $\tz$; two independent mass-squared differences $\ds=m_{2}^{2}-m_{1}^{2}$ and $\dl=m_{3}^{2}-m_{1}^{2}$; and a CP violating phase $\dcp$. The angle $\tx$ and mass-squared difference $\ds$ were measured in the solar neutrino experiments \cite{Bahcall:2004ut, Ahmad:2002jz}, $|\dl|$ was measured in the accelerator neutrino experiment MINOS \cite{Kyoto2012MINOS}, and $\ty$ was measured in the reactor neutrino experiments \cite{An:2012eh, Ahn:2012nd, Abe:2011fz}. However, the sign of $\dl$, the octant of $\tz$ and the value of $\dcp$ are still unknown. 
Currently operating long baseline accelerator neutrino experiments, namely T2K \cite{Itow:2001ee} and \nova \cite{Ayres:2004js}, are expected to measure these unknown quantities. Both the experiments published their data in 2018 and 2019 \cite{Abe:2017vif, Abe:2018wpn, Acero:2019ksn, Abe:2019vii}. According to an analysis of these data sets, the T2K best-fit point is at $\sin^2 \tz=0.53^{+0.03}_{-0.04}$ for both hierarchies, and $\dcp/\pi=-1.89^{+0.70}_{-0.58}$ ($-1.38^{+0.48}_{-0.54}$) for normal (inverted) hierarchy \cite{Abe:2019vii}. On the other hand the \nova best-fit point is at $\sin^2 \tz = 0.56^{+0.04}_{-0.03}$, and $\dcp/\pi=0^{+1.3}_{-0.4}$ for normal hierarchy. Therefore, there is a visible difference between the $\dcp$ values at the \nova and T2K best-fit points. 
A previous result from the \nova collaboration \cite{sanchez_mayly_2018_1286758, NOvA:2018gge} had some mild tension with the T2K data, which has been discussed in ref.~\cite{Nizam:2018got}. In the 2019 analysis with the NO$\nu$A data, the best-fit point is now closer to the T2K best-fit point, but differences still exist as \nova disfavors the T2K best-fit point at $1\, \sigma$ C.L.\ and vice versa. Recently \nova \cite{Himmel:2020} and T2K \cite{Dunne:2020} have published their latest results in the Neutrino, 2020 conference. As per the latest analysis, the best-fit point for  NO$\nu$A (T2K) is $\sin^2\tz= 0.57$ ($0.528$) and $\dcp=0.82 \pi$ ($-1.6\pi$). The tension between the two experiments is even stronger as they exclude each other's allowed region at $1\, \sigma$ C.L. It has also been shown that although both the experiments individually prefer NH over IH, their combined analysis prefers IH over NH \cite{Kelly:2020fkv}. 

Apart from the unknowns in the three flavor neutrino mixing, there are anomalies from the short baseline experiments, which cannot be accounted for in the three flavor oscillation formalism. These anomalies are namely

\begin{enumerate}
\item Reactor anomalies: It implies a deficit of observed $\bar{\nu}_e$ event numbers in different detectors situated at a few meters away from the reactor sources, compared to the predicted number. In particular, the average ratio is $R_{\rm avg}= N_{\rm obs}/N_{\rm pred}=0.927\pm 0.023$ \cite{Mention:2011rk}. Recent updates have changed the predictions slightly, giving an average ratio $R_{\rm avg}=0.938\pm 0.023$ \cite{lasserre}, which is a $2.7\, \sigma$ deviation from unity. However, there is a lack of knowledge of the reactor neutrino fluxes and a detailed study of the forbidden transition in the reactor neutrino spectra may increase the systematic uncertainties to a few percentage. Moreover, there are similar indications of $\nu_e$ disappearance from the SAGE \cite{Abdurashitov:2009tn} and GALLEX \cite{Hampel:1998xg} solar neutrino experiments. A combined analysis of the detected and predicted number of neutrino events from the source gives $R=0.86\pm 0.05$ \cite{Hampel:1997fc, Abdurashitov:1998ne}, another $2.7\, \sigma$ deviation from unity. Both of these deficits of low energy $\bar{\nu}_e$ events can be explained by an oscillation at $\Delta m^2 \geq 1\, {\rm eV}^2$ over very short baseline.
 
\item LSND and MiniBooNE anomalies: The LSND experiment \cite{Aguilar:2001ty} at the Los Alamos National laboratory was designed to observe $\bar{\nu}_\mu\to \bar{\nu}_e$ oscillations over a baseline of 30 m. After 5 years of data taking, it observed $89.7 \pm 22.4 \pm 6.0$ $\bar{\nu}_e$ candidate events over background, providing a $3.8\, \sigma$ evidence of $\bar{\nu}_\mu\to \bar{\nu}_e$ oscillations at $\Delta m^2 = 1\, {\rm eV}^2$ region. Therefore, this result cannot be accommodated in the three flavor scenario. The MiniBooNE experiment \cite{AguilarArevalo:2007it} at the Fermilab was designed to observe $\nu_\mu \to \nu_e$ and $\bar{\nu}_\mu\to \bar{\nu}_e$ oscillations over 540 m baseline using the Booster Neutrino Beam (BNB), a predominantly muon-neutrino beam, peaking at 700 MeV. It observed a $3.4\, \sigma$ signal excess of $\nu_e$ candidate and $2.8\, \sigma$ signal excess of $\bar{\nu}_e$ candidates.
\end{enumerate}

The most common explanation of these anomalies is the existence of one or more ``sterile" neutrino states with mass at or below a few eV range, see ref.~\cite{Abazajian:2012ys} for a comprehensive review.  
The minimal model consists of 3+1 neutrino mixing, dominated by $\nu_e$, $\nu_\mu$ and $\nu_\tau$, with very small perturbative contribution from the new sterile flavor $\nu_x$. The $\nu_x$ mainly consists of a very heavy eigenstate $\nu_4$ with mass $m_4$, such that $m_1,\, m_2,\, m_3\ll m_4$ and $\Delta_{41} = m_{4}^{2}-m_{1}^{2}=[0.1-10]\, {\rm eV}^2$. Recent results from the IceCube experiment constrain the sterile neutrino mass and mixing using atmospheric neutrino fluxes \cite{TheIceCube:2016oqi}. In 2018, however, MiniBooNE has again confirmed a $4.7\, \sigma$ excess of combined $\nu_e$ and $\bar{\nu}_e$ events \cite{Aguilar-Arevalo:2018gpe}. The present significance of the excess from a combined analysis of MiniBooNE and LSND is $6\, \sigma$. Constraints on the existence of sterile neutrino have been discussed in ref.~\cite{Bryman:2019bjg, Boser:2019rta, Miranda:2018buo}, while ref.~\cite{Gupta:2018qsv, Chatla:2018sos, Choubey:2017ppj, Choubey:2017cba, Berryman:2015nua} discuss about the effects of light sterile neutrino on present and future long baseline experiments.

If extra neutrino generations exist as iso-singlet neutral heavy leptons (NHL), in the minimal extension of the standard model, they would not take part in neutrino oscillations, however. The admixture of such leptons in the charged current weak interactions would affect the neutrino oscillation, and the neutrino oscillation would be described by an effective $3\times 3$ non-unitary mixing matrix \cite{Forero:2011pc}. NHL would induce charged lepton flavor violation processes \cite{Forero:2011pc, Bernabeu:1987gr}. If Majorana type, these NHLs will modify the rate of neutrinoless double beta decay \cite{Schechter:1981bd, Rodejohann:2011mu}. The theory of neutrino oscillation in the presence of non-unitarity in the 3-generation scheme and its effect on long baseline accelerator neutrino, have been studied in several references.  Ref.~\cite{Ge:2016xya} has studied the CP violation measurement potential of T2K and future experiment T2HK in presence of non-unitary oscillation, whereas a similar study in the context of DUNE has been done in ref.~ \cite{Escrihuela:2016ube}. In ref.~\cite{Soumya:2018nkw}, physics potential of long baseline experiments with non-unitary mixing has been studied, while ref.~\cite{Fong:2017gke} has made a theoretical study of the non-unitary oscillation at the probability level. In ref.~\cite{Verma:2016nfi}, the CP violation measurement potential of short baseline (SBL) experiments in the presence of non-unitary $\nu_\mu\to\nu_\tau$ oscillation has been studied. All these works were done with simulations for the future experiments, but they did not analyse
already available data, e.g., from the T2K and \nova experiments as we have done in this work. However, in ref.~\cite{Chatterjee:2020kkm, Denton:2020uda} an effort has been made to resolve the tension between \nova and T2K with non-standard NC interaction during propagation and in ref.~\cite{Rahaman:2021leu} the same has been done with CPT conserving Lorentz invariance violation. A global analysis assuming non-unitary hypothesis has been done and limits on non-unitary parameters have been given in ref.~\cite{Escrihuela:2015wra}.

In this paper, we have explored whether the 2019 and 2020 T2K and \nova data can exclude the non-unitary $3\times 3$ mixing, and if not, whether the non-unitary $3\times 3$ mixing hypothesis can lead to better agreement between the T2K and \nova data. A combined analysis with the non-unitary hypothesis has also been done. In Section 2, we have discussed the theory of neutrino oscillation probability in the presence of a non-unitary $3\times 3$ mixing matrix. Details of the simulation method have been discussed in Section 3. In Sections 4 and 5, we have presented our results for the 2019 and 2020 data respectively, and the conclusion has been drawn in Section 6.

\section{Non-unitary oscillation probabilities}
In the standard case of 3 active neutrinos, the flavor basis $\nu_f$ ($f=e, \mu, \tau$) is related to the mass basis $\nu_m$ ($m = 1, 2, 3$) by the relation $\nu_f = U\nu_m$, where $U$ is the unitary PMNS matrix. The Schr\"{o}dinger equation for neutrino propagation in matter can be written as
\begin{equation}
i\frac{d}{dt}\nu_m = (H + U^\dagger \mathcal{A} U)\nu_m\,,
\label{Schroedinger1}
\end{equation}
where $H$ is the Hamiltonian and $\mathcal{A}$ is the matter potential. The solution to the above equation, after propagation over a distance $L$ in matter, is $\nu_m (L) = S_m(L)\nu_m$, where
\begin{equation}
S_m(L) = W e^{-i L \tilde{E}}  W^\dagger = e^{-i LW \tilde{E} W^\dagger}
\label{Smatrix-mass}
\end{equation}
is the $S$-matrix in the mass basis, with $W$ being the transformation matrix between the mass basis $\nu_m$ and the mass basis in matter $\tilde\nu_m$ such that $\nu_m = W\tilde\nu_m$. The energies $\tilde{E} = diag(\tilde{E_1}, \tilde{E_2}, \tilde{E_3})$, where $\tilde{E_i}$ are the eigenvalues in the $\tilde\nu_m$ basis. The $S$-matrix in the flavor basis can be found from the mass basis in eq.~(\ref{Smatrix-mass}) by using the unitary property of the PMNS matrix as
\begin{equation}
S_f(L) = US_m(L)U^\dagger = e^{-iLUW\tilde{E}W^\dagger U^\dagger}.
\label{propagation3*3}
\end{equation}
Flavor change in terms of the $S$-matrix is $\nu_f (L) = S_f(L)\nu_f$. Correspondingly, the oscillation probability from flavor $\alpha$ to flavor $\beta$
can be written as
\begin{equation}
P_{\alpha \beta}=|(S_f(L))_{\beta \alpha}|^2 \,.
\label{probability}
\end{equation}

One can extend this formalism to $n\times n$ unitary mixing matrix $U_{n\times n}$ with $3$ active neutrinos and $(n-3)$ heavy singlet neutrinos \cite{Escrihuela:2015wra}. In that case,
\begin{equation}
U_{n\times n}= \left[ {\begin{array}{cc}
   N_{3\times 3} & Q_{3\times (n-3)}  \\
   V_{(n-3)\times 3} & T_{(n-3)\times (n-3)} 
  \end{array} } \right],
\end{equation}
where $N$ is a $3\times3$ matrix in the light neutrino sector; $Q$ and $V$ depict the coupling parameters of the extra iso-singlet state, expected to be heavy; and $T$ is a $(n-3)\times(n-3)$ matrix in the heavy neutrino sector.
The interaction potential matrix $\mathcal{A}$ can be written as
\begin{equation}
\mathcal{A}_{n\times n}= \left[ {\begin{array}{cc}
   \mathcal{A}_{3\times 3} & 0  \\
   0 & \mathcal{A}_{(n-3)\times (n-3)} 
  \end{array} } \right],
\end{equation}
where, for the usual matter potential, the term $\mathcal{A}_{3\times 3}$ is the same as in eq.~(\ref{Schroedinger1}) and $\mathcal{A}_{(n-3)\times (n-3)} = 0$. An explicit formalism for the potential matrix has been discussed in details in refs.~\cite{Escrihuela:2016ube}.  The Hamiltonian in vacuum in this case can be written as 
\begin{equation}
H_{n\times n}= \left[ {\begin{array}{cc}
   H_{3\times 3} & 0  \\
   0 & H_{(n-3)\times (n-3)} 
  \end{array} } \right],
\end{equation}
where $H_{3\times 3}$ is the Hamiltonian in vacuum from eq.~(\ref{Schroedinger1}). Similarly, the neutrino flavor vector can be rewritten as 
\begin{equation}
\nu_{f_{n}}= \left[ {\begin{array}{c}
   \nu_{f_{3}}  \\
   \nu_{f_{(n-3)}}
  \end{array} } \right],
  \label{n-neutrinos}
\end{equation}
where we split the flavor vector in active $\nu_{f_{3\times 1}}$ and heavy $\nu_{f_{(n-3)\times 1}}$ parts. The $W$ matrix similar to that in eq.~(\ref{Smatrix-mass}) can be written as
\begin{equation}
W_{n\times n}= \left[ {\begin{array}{cc}
   W_{3\times 3} & W_{3\times (n-3)}  \\
   W_{(n-3)\times 3} & W_{(n-3)\times (n-3)} 
  \end{array} } \right],
\end{equation}
and the $\tilde{E}$ matrix from eq.~(\ref{Smatrix-mass}) can be written as
\begin{equation}
\tilde{E}_{n\times n}= \left[ {\begin{array}{cc}
   \tilde{E}_{3\times 3} & 0  \\
   0 & \tilde{E}_{(n-3)\times (n-3)} 
  \end{array} } \right].
\end{equation}

The Hamiltonian in the presence of matter potential, following eq.~(\ref{Schroedinger1}), is
\begin{eqnarray}
H_{m_{n\times n}} &=& H_{n\times n} + U_{n\times n}^\dagger \mathcal{A}_{n\times n} U_{n\times n} \\
&=& \left[ {\begin{array}{cc}
   H_{3\times 3}+N_{3\times 3}^\dagger   \mathcal{A}_{3\times 3} N_{3\times 3} \,+ & 
   N_{3\times 3}^\dagger \mathcal{A}_{3\times 3} Q_{3\times (n-3)} \,+ \\
   V_{(n-3)\times 3}^\dagger \mathcal{A}_{(n-3)\times (n-3)}V_{(n-3)\times 3} & \, V_{(n-3)\times 3}^\dagger \mathcal{A}_{(n-3)\times (n-3)}T_{(n-3)\times (n-3)}  \\
   \\
    Q_{3\times (n-3)}^\dagger \mathcal{A}_{3\times 3} N_{3\times 3} \, + &  
    H_{(n-3)\times (n-3)}+Q_{3\times (n-3)}^\dagger \mathcal{A}_{3\times 3} Q_{3\times (n-3)}\, + \\
    T_{(n-3)\times (n-3)}^\dagger \mathcal{A}_{(n-3)\times (n-3)}V_{(n-3)\times 3} & T_{(n-3)\times (n-3)}^\dagger \mathcal{A}_{(n-3)\times (n-3)}T_{(n-3)\times (n-3)}   
  \end{array} } \right]. \nonumber
\end{eqnarray}
Similarly, the $S$-matrix in the mass basis, following eq.~(\ref{Smatrix-mass}), is
\begin{eqnarray}
S_{m_{n\times n}}(L) &=& W_{n\times n}e^{-iL\tilde{E}}W_{n\times n}^\dagger \\
&=&\left[ {\begin{array}{cc}
   W_{3\times 3}e^{-iL\tilde{E}_{3\times3}}W_{3\times 3}^\dagger\, + &   
   W_{3\times 3}e^{-iL\tilde{E}_{3\times3}}W_{(n-3)\times 3}^\dagger\, + \\
   W_{3\times (n-3)} e^{-iL\tilde{E}_{(n-3)\times(n-3)}} W_{3\times (n-3)}^\dagger &  W_{3\times (n-3)} e^{-iL\tilde{E}_{(n-3)\times(n-3)}} W_{(n-3)\times (n-3)}^\dagger \\
   \\
    W_{(n-3)\times 3}e^{-iL\tilde{E}_{3\times3}}W_{3\times 3}^\dagger\, + & 
    W_{(n-3)\times 3}e^{-iL\tilde{E}_{3\times3}}W_{(n-3)\times 3}^\dagger\, + \\
    W_{(n-3)\times (n-3)}e^{-iL\tilde{E}_{(n-3)\times(n-3)}} W_{3\times (n-3)}^\dagger & 
    W_{(n-3)\times (n-3)}e^{-iL\tilde{E}_{(n-3)\times(n-3)}} W_{(n-3)\times (n-3)}^\dagger
  \end{array} } \right]. \nonumber
  \end{eqnarray}
The masses of the heavy right handed neutrinos are expected to be several orders of magnitude larger than the masses of the active neutrinos and the potential terms. Under this condition the elements $W_{3 \times (n-3)}$ and $W_{(n-3)\times 3}$ satisfy conditions that they are $\ll 1$, following the typical see-saw mechanisms \cite{Grimus:2000vj, Hettmansperger:2011bt}.
Therefore, we can neglect these terms and write the $S$-matrix as
\begin{eqnarray}
S_{m_{n\times n}}(L)&\approx & \left[ {\begin{array}{cc}
   W_{3\times 3}e^{-iL\tilde{E}_{3\times3}}W_{3\times 3}^\dagger &   0\\
   0 &  W_{(n-3)\times (n-3)} e^{-iL\tilde{E}_{(n-3)\times(n-3)}} W_{(n-3)\times (n-3)}^\dagger
      \end{array} } \right] \nonumber \\
      &\equiv &  \left[ {\begin{array}{cc}
   S_{m_{3\times 3}} & 0\\
   0 &  S_{m_{(n-3)\times(n-3)}}
      \end{array} } \right].
\end{eqnarray}
The $S$-matrix in the flavor basis is written, following eq.~(\ref{propagation3*3}), as
\begin{eqnarray}
S_{f_{n\times n}}(L) &=& U_{n\times n} S_{m_{n\times n}}(L)U_{n\times n}^\dagger \nonumber \\ &=& \left[ {\begin{array}{cc}
N S_{m_{3\times 3}} N^\dagger + Q S_{m_{(n-3)\times(n-3)}} Q^\dagger &  
N S_{m_{3\times 3}} V^\dagger + Q S_{m_{(n-3)\times(n-3)}} T^\dagger\\
V S_{m_{3\times 3}} N^\dagger + T S_{m_{(n-3)\times(n-3)}} Q^\dagger &  
V S_{m_{3\times 3}} V^\dagger + T S_{m_{(n-3)\times(n-3)}} T^\dagger 
\end{array} } \right] .\nonumber \\ 
\end{eqnarray}

In case of 3 active neutrinos the flavor changes, following eq.~(\ref{n-neutrinos}), as
\begin{eqnarray}
\nu_{f_{3}}(L) &=& (N S_{m_{3\times 3}} N^\dagger + Q S_{m_{(n-3)\times(n-3)}} Q^\dagger )\nu_{f_3} \nonumber \\
&& + (N S_{m_{3\times 3}} V^\dagger + Q S_{m_{(n-3)\times(n-3)}} T^\dagger ) \nu_{f_{(n-3)}}.
\end{eqnarray}
Here $S_{m_{3\times 3}}$ is the same as in eq.~(\ref{Smatrix-mass}). With a similar see-saw mechanism argument due to the connection between the mass and flavor bases, the elements of $Q$ and $V$ are much lower compared to the terms containing $N$ and $T$ matrices \cite{Escrihuela:2015wra, Hettmansperger:2011bt}. As a good approximation, we can therefore eliminate terms with $Q$ and $V$ matrices and write the non-unitary solution for 3-active neutrino flavor states as
\begin{equation}
\nu_{f_{3}}(L) \approx N e^{-iLW_{3\times 3}\tilde{E}_{3\times 3}W_{3\times 3}^{\dagger}} N^\dagger \nu_{f_3} \equiv S_f(L)\nu_{f_3}.
\end{equation}
The corresponding oscillation probability can be calculated using eq.~(\ref{probability}). 

We use the following parametrization of the non-unitary mixing matrix $N$ \cite{Escrihuela:2015wra}
\begin{equation}
N=N_{NP}U_{3\times 3}= \left[ {\begin{array}{ccc}
   \alpha_{00} & 0 & 0 \\
   \alpha_{10} & \alpha_{11} & 0 \\
   \alpha_{20} & \alpha_{21} & \alpha_{22}
  \end{array} } \right] U_{\rm PMNS} \,,
\end{equation}
where the diagonal term(s) must deviate from unity and/or the off-diagonal term(s) deviate from zero to allow non-unitarity effect. The oscillation experiments can probe non-unitarity only if the $\alpha$ parameters vary at least at the percent scale \cite{Forero:2011pc, Escrihuela:2015wra, Blennow:2016jkn}. There exists severe constraint on the parameter $|\alpha_{10}| < 10^{-5}$ from non-observation of the $\mu\to e\gamma$ decay \cite{Blennow:2016jkn}. However, this bound can be relaxed in certain neutrino mass-generation models involving inverse or linear see-saw mechanism \cite{Forero:2011pc}.
In this paper, to calculate the probabilities, we have kept the $\alpha$ parameters fixed at their $3\,\sigma$ boundary values \cite{Escrihuela:2016ube}: 
\begin{eqnarray}
\alpha_{00} > 0.93 \,;\, \alpha_{11} > 0.95 \,;\, \alpha_{22} > 0.61 \nonumber \\ 
|\alpha_{10}| < 3.6\times 10^{-2} \,;\, |\alpha_{20}| < 1.3\times 10^{-1} 
\,;\, |\alpha_{21}| < 2.1\times 10^{-2}
\label{eq:nubounds}
\end{eqnarray}
The standard unitary parameter values have been fixed at their best-fit values \cite{Esteban:2018azc}. 

\begin{figure}[htbp]
\centering
\includegraphics[width=0.6\textwidth]{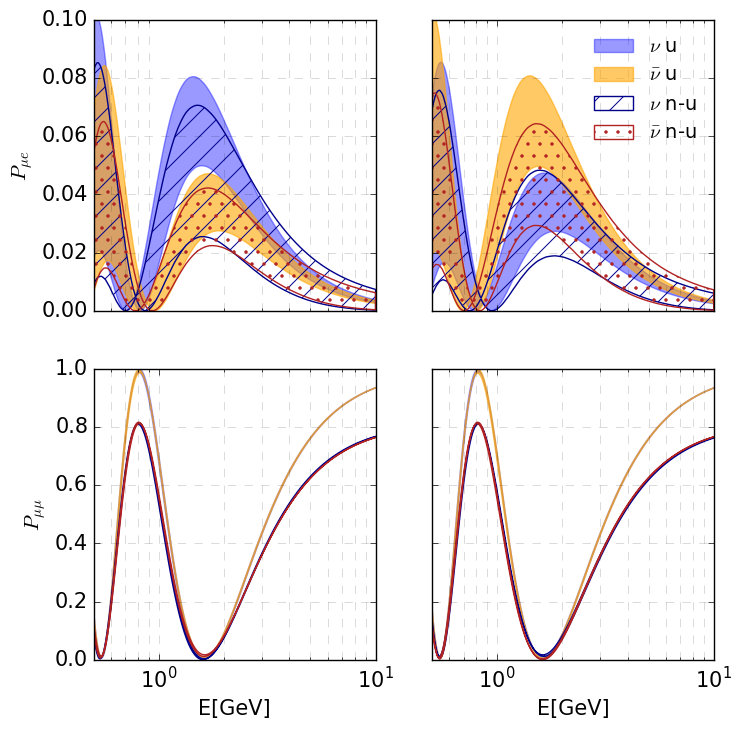}
\caption{\footnotesize{Comparison between unitary and non-unitary $3\times 3$ mixing of $\nu_\mu \to \nu_e$ ($\nu_\mu \to \nu_\mu$) probabilities in the upper (lower) panel for the \nova experiment (810 km baseline). Left (right) panel is for normal hierarchy (inverted hierarchy). The unitary CP-violating phase $\dcp$ has been varied in its total range $[-180^\circ:180^\circ]$. Non-unitary complex phase $\phi_{10}=0$. Non-unitary parameters are fixed at their boundary values in eq.~(\ref{eq:nubounds}) taken from ref.~\cite{Escrihuela:2016ube}.}}
\label{prob1}
\end{figure}
\begin{figure}[htbp]
\centering
\includegraphics[width=0.6\textwidth]{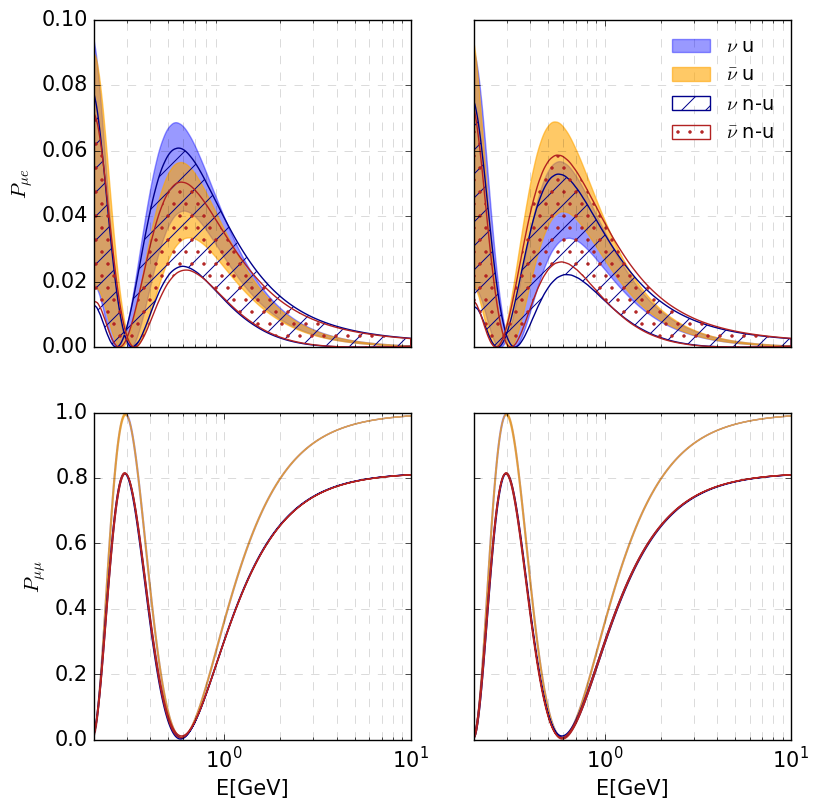}
\caption{\footnotesize{Same as Fig.~\ref{prob1} but for the T2K experiment (295 km baseline).}}
\label{prob2}
\end{figure}

For the standard unitary case, the $\nu_\mu \to \nu_e$ oscillation probability can be written as \cite{Cervera:2000kp}
\begin{eqnarray}
  \pme &\simeq& \sin^2 2 \ty \sin^2 \tz\frac{\sin^2\dhat(1-\ahat)}{(1-\ahat)^2}\nonumber\\
  &+& \alpha \cos \ty \sin2\tx \sin 2\ty \sin 2\tz \cos(\dhat+\dcp)
 \frac{\sin\dhat \ahat}{\ahat}
  \frac{\sin \dhat(1-\ahat)}{1-\ahat},
  \label{pme}
   \end{eqnarray}
where $\alpha=\frac{\ds}{\dl}$, $\dhat=\frac{\dl L}{4E}$ and $\ahat=\frac{A}{\dl}$. $A$ is the Wolfenstein matter term \cite{msw1}, given by $A=2\sqrt{2}G_FN_eE$, where $E$ is the neutrino beam energy and $L$ is the length of the baseline. Anti-neutrino oscillation probability $\pmebar$ can be obtained by changing the sign of $A$ and $\dcp$ in eq.~\ref{pme}. The oscillation probability mainly depends on hierarchy (sign of $\dl$), octant of $\tz$ and $\dcp$.

From eq.~(\ref{pme}), we can see that for the standard unitary case, the oscillation probability $P_{\mu e}$ ($P_{\bar{\mu}\bar{e}}$) gets a double boost (double suppression) when the hierarchy is NH and $\dcp$ is in the lower half plane (LHP). Similarly $P_{\mu e}$ ($P_{\bar{\mu}\bar{e}}$) gets a double suppression (double boost) when the hierarchy is IH and $\dcp$ is in the upper half plane (UHP). Therefore $P_{\mu e}$ ($P_{\bar{\mu}\bar{e}}$) is maximum (minimum) for NH and $\dcp$ in the LHP and minimum (maximum) for IH and $\dcp$ in the UHP. 

\begin{figure}[htbp]
\centering
\includegraphics[width=0.6\textwidth]{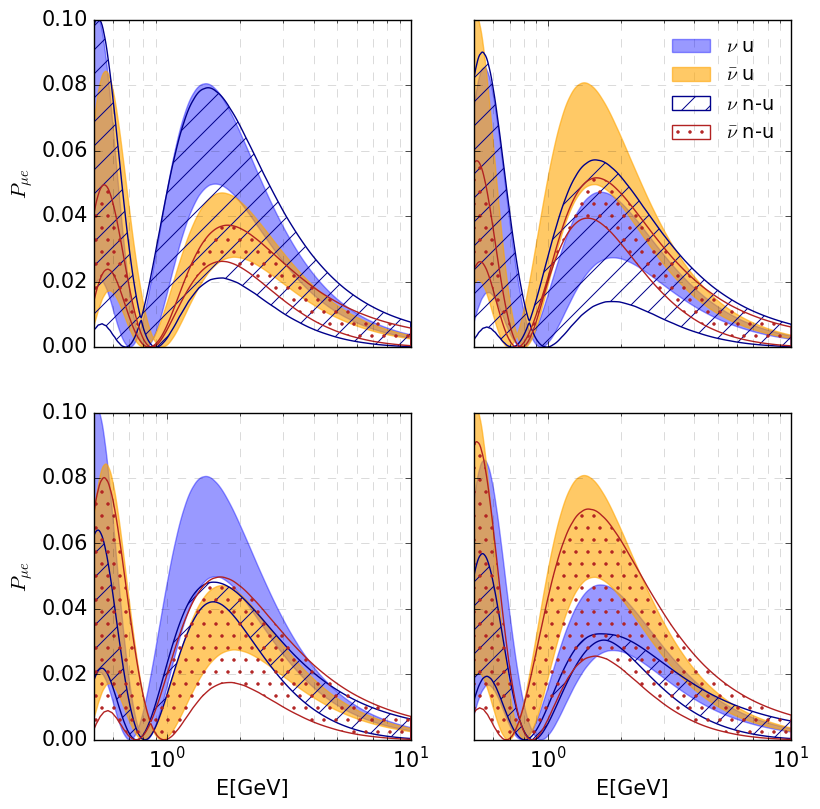}
\caption{\footnotesize{Comparison between the unitary and non-unitary  $\nu_\mu \to \nu_e$ probabilities for $\phi_{10}=\pi/2$ ($-\pi/2$) in the upper (lower) panel for the \nova experiment. Left (right) panel is for Normal hierarchy (inverted hierarchy). The unitary CP violating phase $\dcp$ has been varied in its total range $[-180^\circ:180^\circ]$. The non-unitary parameters are fixed at their boundary values in eq.~(\ref{eq:nubounds}) taken from ref.~\cite{Escrihuela:2016ube}.}}
\label{prob3}
\end{figure}

\begin{figure}[htbp]
\centering
\includegraphics[width=0.6\textwidth]{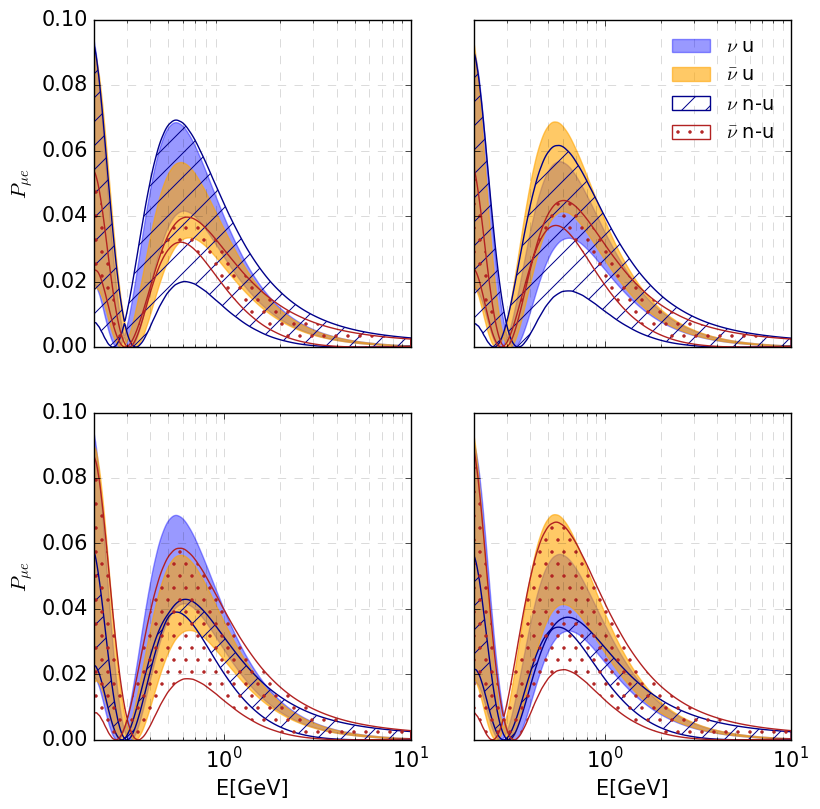}
\caption{\footnotesize{Same as Fig.~\ref{prob3} but for the T2K experiment.}}
\label{prob4}
\end{figure}

In the case of non-unitary mixing, an analytic expression for $\nu_\mu \to \nu_e$ oscillation with matter effect is not available in literature and it is difficult to calculate them. However, approximate analytic oscillation probability in vacuum with non-unitary mixing is given in different articles, see, e.g.,  \cite{Escrihuela:2015wra}. From the expression of vacuum oscillation probability with non-unitary mixing, we can see that the parameters $\alpha_{ij}$'s, where $i,\, j=0,\, 1, \, 2$, are multiplied with different unitary parameters. Sinces $\alpha_{ij}<1$, we can expect them to reduce the oscillation probability in case of non-unitary mixing. This feature is visible in figures \ref{prob1}-\ref{prob4}.

In Fig.~\ref{prob1}, we have shown the comparison of the $\nu_\mu \to \nu_e$ oscillation probabilities and $\nu_\mu \to \nu_\mu$ survival probabilities between the unitary (labeled {\bf u}) and non-unitary (labeled {\bf n-u}) $3\times 3$ mixing, for both the normal hierarchy (NH) and inverted hierarchy (IH) in the case of the \nova experiment baseline. The unitary CP violating phase $\dcp$ has been kept as a floating parameter. The phases associated with the non-diagonal elements of the $\alpha$ matrix are zero. Similar comparison has been done for the T2K experiment as well in Fig.~\ref{prob2}. In the case of unitary mixing, the upper (lower) boundary for $P_{\mu e}$ denotes $\dcp=-90^\circ$ ($90^\circ$). For the anti-neutrino oscillation, this characteristic is just the opposite. The probability for all other $\dcp$ values fall in between. In both figures, the disappearance plots in lower panel look similar for both the hierarchies. This is because the $\nu_\mu$ survival probability does not get affected by matter and hence is not sensitive to hierarchy.

From both Figs.~\ref{prob1} and \ref{prob2} bottom panels, it is obvious that for $\nu_\mu$ disappearance channels the survival probability for unitary $3\times 3$ mixing matrix can be differentiated from the non-unitary mixing at 
energies away from the oscillation maxima and flux peak energy of \nova (2.0~GeV flux peak energy) and T2K (0.7~GeV flux peak energy) experiments. It is also observable that there are overlaps between the unitary and non-unitary probabilities for both $\nu_e$ and $\bar{\nu}_e$ appearances, and therefore, the appearance channels reduce sensitivity to differentiate between the unitary and non-unitary mixing in these experiments.

In Figs.~\ref{prob3} and \ref{prob4}, we have compared between the oscillation probabilities of the unitary and non-unitary case for $\phi_{10}=\pi/2$ and $-\pi/2$, respectively, where $\phi_{10}$ is the phase associated with $\alpha_{10}=|\alpha_{10}|e^{i\phi_{10}}$. For both NH and IH, the oscillation probabilities in the neutrino channel, can be mimicked by the non-unitary oscillation probability with $\phi_{10}=\pi/2$ and it holds for both \nova and T2K. For this value of $\phi_{10}$, the discrimination between the unitary and non-unitary cases is better in the anti-neutrino channel. However, for $\phi_{10}=-\pi/2$ and for both the experiments, the unitary oscillation probabilities in the anti-neutrino channel can be mimicked by the non-unitary probability. The neutrino channel has better discrimination capability between the unitary and non-unitary cases for both the experiments in this case. Moreover, for T2K, the unitary oscillation probabilities in the neutrino (anti-neutrino) channel, with NH and almost for the total range of $\dcp$, can be mimicked by the non-unitary probabilities with IH and $\phi_{10}=\pi/2$ ($-\pi/2$). It is clear from this discussion that data from the \nova and T2K experiments may not constrain the non-unitarity significantly.

\begin{figure}[htbp]
\centering
\includegraphics[width=0.6\textwidth]{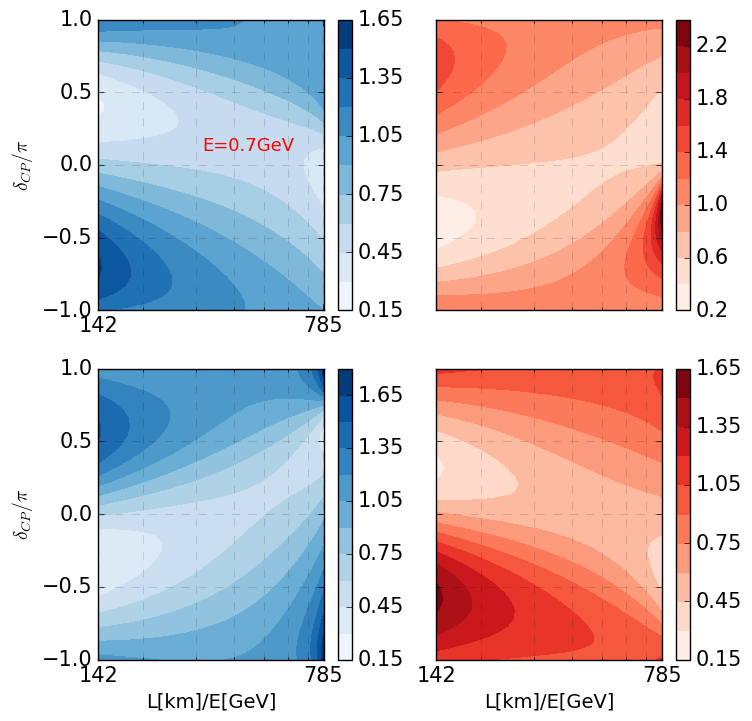}
\caption{\footnotesize{Ratio of the non-unitary to unitary oscillation probabilities as a function of $L/E$ and $\dcp$. The reference energy $E=0.7$ GeV has been fixed to the T2K flux peak energy. For this peak energy and T2K baseline, $L/E = 421$. The upper (lower) panel shows the ratio for NH (IH), the left (right) panel shows it for neutrino (anti-neutrino). The non-unitary parameters are fixed to their boundary values in eq.~(\ref{eq:nubounds}) taken from ref.~\cite{Escrihuela:2016ube} and we have set $\phi_{10}=0$.}}
\label{prob5}
\end{figure}

\begin{figure}[htbp]
\centering
\includegraphics[width=0.6\textwidth]{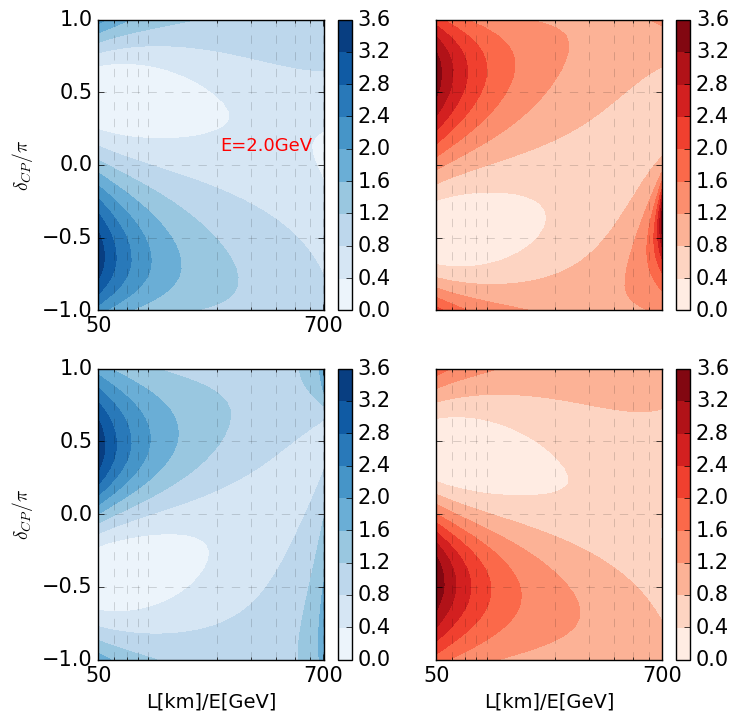}
\caption{\footnotesize{Same as Fig.~\ref{prob5} but the reference energy $E=2.0$ GeV has been fixed to the \nova flux peak energy. For this peak energy and \nova baseline, $L/E = 405$.}}
\label{prob6}
\end{figure}

\begin{figure}[htbp]
\centering
\includegraphics[width=0.6\textwidth]{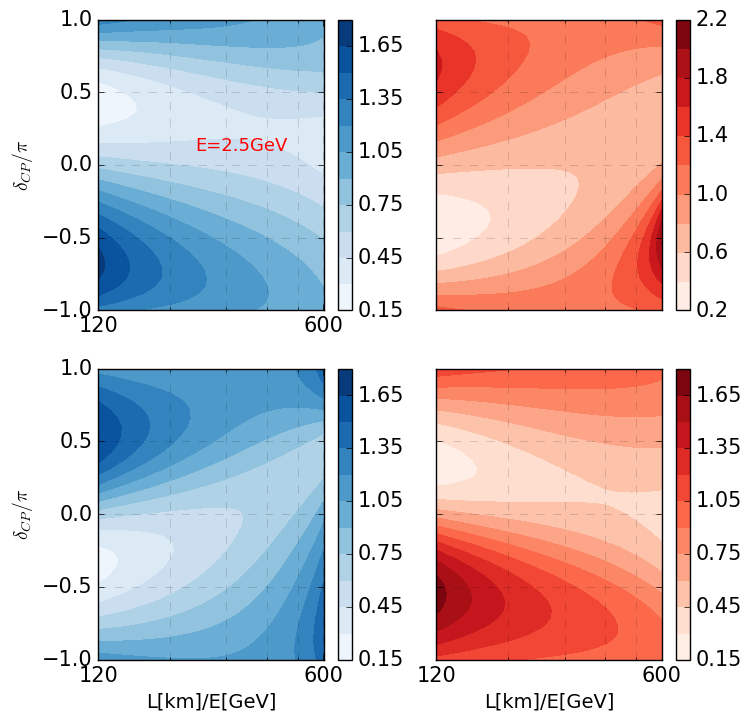}
\caption{\footnotesize{Same as Fig.~\ref{prob5} but the reference energy $E=2.5$ GeV has been fixed to the DUNE flux peak energy. For this peak energy and DUNE baseline, $L/E = 520$.}}
\label{prob7}
\end{figure}

\begin{figure}[htbp]
\centering
\includegraphics[width=0.6\textwidth]{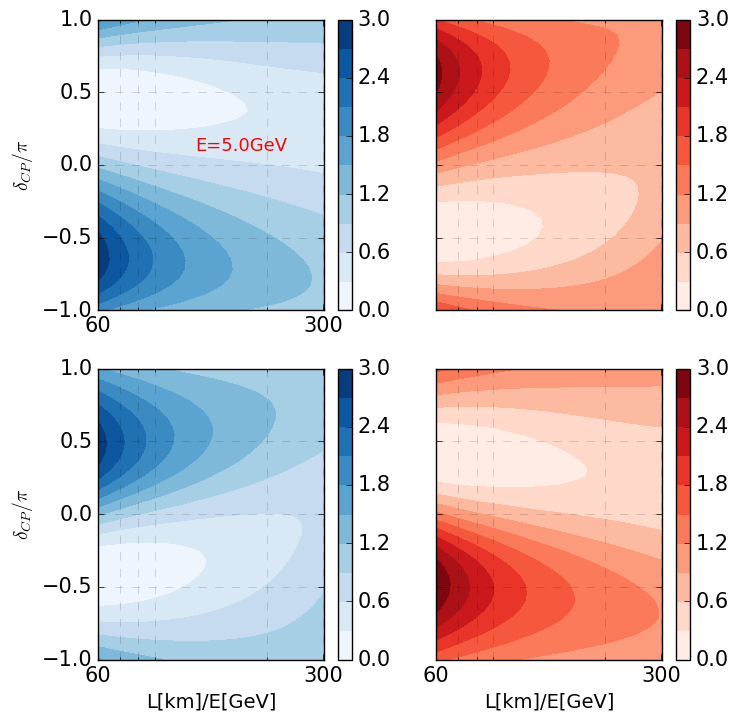}
\caption{\footnotesize{Same as Fig.~\ref{prob5} but the reference energy $E=5.0$ GeV has been fixed to the flux peak energy of an arbitrary future experiment.}}
\label{prob8}
\end{figure}

In Figs.~\ref{prob5}-\ref{prob8}, we have shown the ratio of the non-unitary to unitary \red{$\nu_\mu \to \nu_e$} oscillation probabilities as a function of $L/E$ and $\dcp$, where $L$ is the baseline of an experiment and $E$ is the energy of the neutrino beam. To do this, we fixed $E$ to four values, namely 0.7 GeV, 2 GeV, 2.5 GeV and 5 GeV and for each $E$ we varied $L$ from 100 km to no more than 1500 km to obtain different $L/E$ values. The non-unitary parameters are fixed at their boundary values as before and $\phi_{10}=0$. The first three energy values are respectively for the NO$\nu$A, T2K and DUNE flux peak points. The farther away the ratio is from $1$, the better the potential to differentiate between the two mixing scenarios. 
In general the sensitivity to non-unitarity is higher at the two opposing corners of the plots, e.g., for $\dcp < 0$ with smaller values of $L/E$ or $\dcp > 0$ with larger values of $L/E$ for the neutrino channel in NH. The opposite is true for the antineutrino channels and so on.
Hence, it is clear from the figures that \nova and T2K, along with DUNE, have discrimination capability between the non-unitary and unitary cases for a very small range of $\dcp$. If a future experiment can be built with smaller baseline and larger flux peak energy, it will be possible to differentiate the non-unitary mixing from the unitary one better. 

\section{Simulation details}
\label{simulations}
The T2K experiment \cite{Itow:2001ee} uses the $\nu_\mu$ beam from the J-PARC accelerator at Tokai and the water Cerenkov detector at Super-Kamiokande, which is 295 km away from the source. The detector is situated $2.5^\circ$ off-axis. The flux peaks at $0.7$ GeV, which is also close to the first oscillation maximum.  T2K started taking data in 2009 and until 2019 release of results, these \cite{Abe:2017vif, Abe:2018wpn, Abe:2019vii} correspond to $14.9 \times 10^{20}$ $(16.4\times 10^{20})$ protons on target (POT) in neutrino (anti-neutrino) mode. The \nova detector \cite{Ayres:2004js} is a 14 kt totally active scintillator detector (TASD), placed 810 km away from the neutrino source at the Fermilab and it is situated at $0.8^\circ$ off-axis of the NuMI beam. The flux peaks at $2$ GeV, close to the oscillation maxima at 1.4 GeV for NH and at 1.8 GeV for IH. \nova started taking data in 2014 and took data until 2019 release \cite{Acero:2019ksn}, these correspond to $8.85 \times 10^{20}$ $(6.9 \times 10^{20})$ POTs, for neutrino (anti-neutrino) mode.

\begin{figure}[htbp]
\centering
\includegraphics[width=0.85\textwidth]{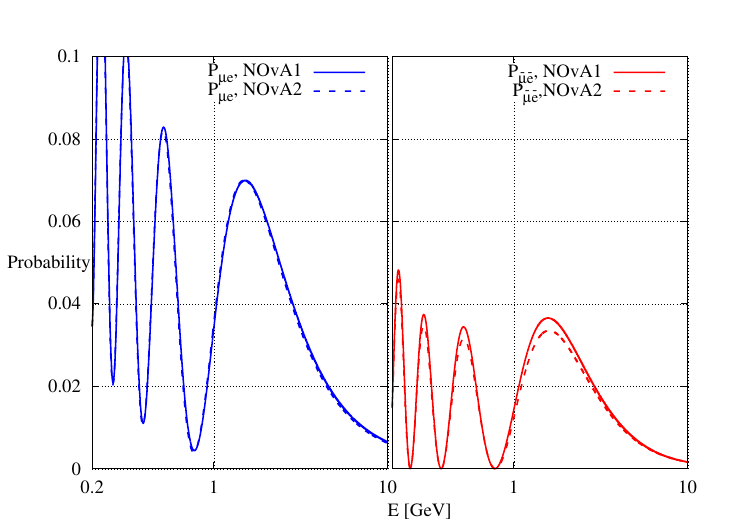}
\includegraphics[width=0.85\textwidth]{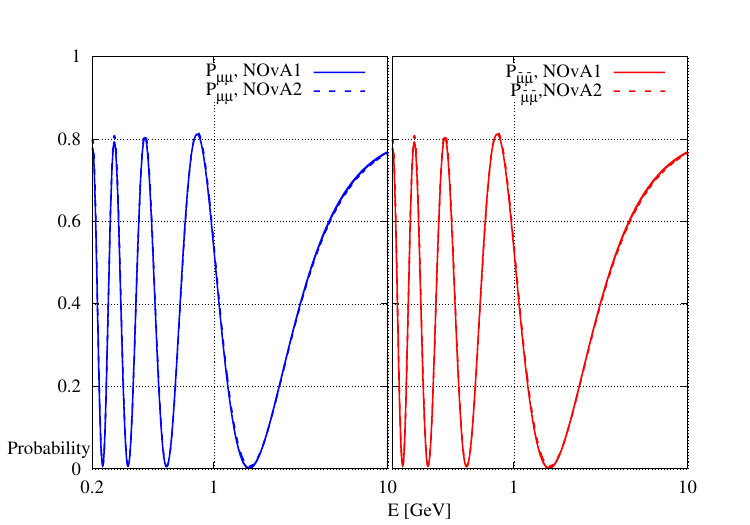}
\caption{\footnotesize{Comparison of probabilities for \nova with NH and different sets of values for $\alpha_{20}$, $\alpha_{21}$ and $\alpha_{22}$. For NOvA1 (NOvA2), $\alpha_{00}=0.93$; $\alpha_{11}=0.95$;
$|\alpha_{10}=3.6\times 10^{-2}|$; $\alpha_{20}=1.3\times 10^{-1}$ ($0$); $\alpha_{21}=2.1\times 10^{-2}$ ($0$); $\alpha_{22}=0.61$ ($1$). The standard oscillation parameters are fixed at their NH best-fit values, taken from ref.~\cite{nufit, Esteban:2018azc}. The left (right) panel shows the probabilities for neutrino(anti-neutrino) channel and the
upper (lower) panel shows probabilities for appearance (disappearance).}}
\label{prob_comp_alpha}
\end{figure}

To analyse the T2K and \nova data, we have taken the solar neutrino parameters $\ds$ and $\sin^2 \tx$ to be fixed at $7.50 \times 10^{-5}\, {\rm eV}^2$ and $0.30$, respectively. For the reactor neutrino angle, $\sin^2 2\ty$ has been varied in its $3~\sigma$ range around the central value of $0.084$ with $3.5 \%$ uncertainty \cite{Ochoa:2018fjq}. For the atmospheric mixing angle, $\sin^2 \tz$ has been varied in the $3\, \sigma$ range $[0.40:0.63]$ \cite{nufit, Esteban:2018azc}. The atmospheric effective mass squared difference $\Delta m^2_{\rm eff}$ has been varied in the MINOS $3~\sigma$ range around the best-fit value of $2.32 \times 10^{-3}\, {\rm eV}^2$ \cite{Adamson:2011ig}. The effective mass-squared difference $\Delta m^{2}_{\rm eff}$ is related with $\dl$ by the following relation \cite{Nunokawa:2005nx}: 
\begin{equation}
\Delta m^{2}_{\rm eff}= \sin^2 \tz \dl + \cos^2 \tx \Delta_{32}+\cos \dcp \sin 2\tx \sin \ty \tan \tx \ds.
\end{equation}
Among the non-unitary parameters, $\alpha_{00},\, \alpha_{11}$, $|\alpha_{10}|$ and $\phi_{10}$ have been varied, while the other non-unitary parameters have been kept constant at their boundary values. This is because only these non-unitary parameters affect appreciably the $\nu_\mu \to \nu_e$ oscillation probability and the $\nu_\mu \to \nu_\mu$ survival probability. The choice of the values of $\alpha_{20}$, $\alpha_{21}$ and $\alpha_{22}$ is justified in fig.~\ref{prob_comp_alpha}.
In this plot, we have shown the comparison between probabilities for \nova when all non-unitary parameters are fixed at their boundary values (denoted by NOvA1) and when $\alpha_{00}$, $\alpha_{11}$, and $\alpha_{10}$ are
at their boundary values, but other non-unitary parameters are fixed at their unitary values (denoted by NOvA2). We can see that the difference between these two probabilities are negligible. The unitary parameters are fixed at
their present best-fit values taken from ref.~\cite{nufit, Esteban:2018azc} and the hierarchy is NH. We further calculated the $\chi^2$ between these two parameter sets for neutrino and anti-neutrino run in \nova with the latest POT. The $\chi^2$
value between these two sets of parameter with the latest POT in \nova happens to be $0.2$. Therefore, it is safe to say that the choice of $\alpha_{20}$, $\alpha_{21}$ and $\alpha_{22}$ do not have any significant effect on the present 
accelerator neutrino data
and we can fix them at their boundary values without any controversy.

We have used GLoBES \cite{Huber:2004ka, Huber:2007ji} to calculate the binned theoretical event rates as a function of the test values of the oscillation parameters. The energy dependent efficiencies of the detector for both signal and 
background events have been fixed according to the
expected event rates plots as a function of energy, given in \cite{Acero:2019ksn, Abe:2017vif, Abe:2018wpn, Abe:2019vii}. For $\nu_e$ appearance data, we considered backgrounds from $\nu_\mu$ CC interactions, beam contamination and NC interactions. For $\nu_\mu$ disappearance data, the backgrounds were from $\nu_e$ CC interactions, beam contamination and NC interactions. Automatic bin based energy smearing for generated theoretical events has been implemented in the same way as described in the GLoBES manual \cite{Huber:2004ka, Huber:2007ji}.
For this purpose, we used a Gaussian smearing function
\begin{equation}
R^c (E,E^\prime)=\frac{1}{\sqrt{2\pi}}e^{-\frac{(E-E^\prime)^2}{2\sigma^2(E)}},
\end{equation}
where $E^\prime$ is the reconstructed energy. The energy resolution function is given by 
\begin{equation}
\sigma(E)=\alpha E+\beta \sqrt{E}+\gamma,
\label{res}
\end{equation}
where $\alpha=0$, $\beta=0.075$, $\gamma=0.05$ for T2K. For NO$\nu$A, however, we used $\alpha=0$, $\beta=0.085$ (0.06), and $\gamma=0$ for $\nu_e$ ($\nu_\mu$) events.
For NC events, in NO$\nu$A, we used migration matrices as discussed in \cite{Agarwalla:2012bv}. The similar energy smearing techniques have been used in refs.\cite{Prakash:2013dua, Bharti:2018eyj, Nizam:2018got}.
 
The experimental event rates have been taken from \nova \cite{Acero:2019ksn} and T2K \cite{Abe:2017vif, Abe:2018wpn, Abe:2019vii} collaboration papers. The $\chi^2$ between the theory and experiments have been calculated for the appearance and disappearance channels for both the neutrino and anti-neutrino runs of both the experiments. To generate $\chi^2$, we have used 30 million data points in the parameter ranges stated above. We have used the Poissonian $\chi^2$ formula:
\begin{eqnarray}
%
\chi^2 &=& 2\sum_i \left\{
(1+z) N_i^{\rm th} - N_i^{\rm exp} + N_i^{\rm exp} 
\ln\left[ \frac{N_i^{\rm exp}}{(1+z) N_i^{\rm th}} \right]
\right\} + 2 \sum_j (1+z) N_j^{\rm th} + z^2 \,\, \nonumber \\
\label{poisionian}
\end{eqnarray}
where $i$ stands for the bins for which $N_i^{\rm exp}\neq 0$ and  $j$ stands for the bins for which $N_j^{\rm exp} = 0$. The parameter $z$ defines the additional systematic uncertainties. For each of the two experiments, we have included systematic uncertainties of $10\%$, using the pull method. We have varied the pull parameters in their $3\sigma$ range and have marginalized over it to calculate the $\chi^{2}_{m}$ as a function of the test values of the oscillation parameters and mass hierarchies. For a particular experiment, the total $\chi^2$ is calculated by 
\begin{eqnarray}
\chi^2({\rm tot})&=&\chi^{2}_{m}(\nu\, {\rm app})+ \chi^{2}_{m}(\anu\, {\rm app})+ \chi^{2}_{m}(\nu\, {\rm disapp})\nonumber \\
&&+ \chi^{2}_{m}(\anu\, {\rm disapp})+\chi^2({\rm prior})
\end{eqnarray}
During the calculation of $\chi^2({\rm tot})$, we have to keep in mind that the test values of the oscillation parameters are same for all the individual $\chi_{m}^{2}$s. The $\chi^2({\rm tot})$ is a function of the test values of the oscillation parameters and hierarchies. The definitions of the $\chi^2({\rm prior})$ and its significance have been discussed in details in ref.~\cite{Gandhi:2007td}. In our analysis, we have used priors to $\sin^2 2\ty$, $\sin^2 \tz$, and $|\Delta m^{2}_{\rm eff}|$. Then, we found out the minimum $\chi^{2}({\rm tot})$ and subtracted it from the $\chi^{2}({\rm tot})$ values to calculate the $\dchsq$ as a function of the oscillation parameters. 

To do a combined analysis of the \nova and T2K data, we define the total $\chi^2$ as:
\begin{eqnarray}
\chi^2({\rm tot})&=&\chi^{2}_{m}({\rm NO}\nu{\rm A}\, \nu\, {\rm app})+ \chi^{2}_{m}({\rm NO}\nu{\rm A}\, \anu\, {\rm app})+\chi^{2}_{m}({\rm T2K}\, \nu\, {\rm app})+ \chi^{2}_{m}({\rm T2K}\, \anu\, {\rm app}) \nonumber \\
&&+ \chi^{2}_{m}({\rm NO}\nu{\rm A}\, \nu\, {\rm disapp}) + \chi^{2}_{m}({\rm NO}\nu{\rm A}\, \anu\, {\rm disapp}) \nonumber \\
&&+\chi^{2}_{m}({\rm T2K}\, \nu\, {\rm disapp})+ \chi^{2}_{m}({\rm T2K}\, \anu\, {\rm disapp})+\chi^2({\rm prior})
\end{eqnarray}
Just like the separate analysis, priors have been added for the $\sin^2 2\ty$, $\sin^2 \tz$ and $|\Delta m^{2}_{\rm eff}|$. The $\dchsq$ has been calculated as before.

In this work, we did not simulate the near detector in detail. The effect of the near
detector is included in both T2K and \nova as errors in the systematic uncertainties. This
procedure is the common approach taken in the literature. This approximation is valid in some specific regimes in the presence of extra heavy neutrinos, see ref.~\cite{Blennow:2016jkn} for the discussion. We also remark that in order for the bounds on non-unitarity in any regime from near detectors measurements to be competitive, it is necessary to have a very good knowledge of the flux arriving at the near detectors (See a more detailed discussion on the impact of the flux uncertainties in non-unitary and near detectors in~\cite{Miranda:2018yym}). Because the uncertainties in the near detector flux for both T2K and NO$\nu$A are much larger than the present bounds on those parameters, our sensitivity comes entirely from the non-unitary effects of the propagation of the three (active) neutrinos. This implies that our bounds are valid for
heavy neutrinos whose oscillations are averaged out or not produced because their masses are heavier than the experimental energy. Therefore, the systematic-like near detector effect approximation that we have used gives conservative bounds.

\section{Analysis of 2019 data}
In this section, we discuss the analysis of the 2019 T2K and \nova data individually with the hypothesis of non-unitary $3\times 3$ mixing matrix. We have also done a combined analysis of the T2K and \nova 2019 data with the non-unitary hypothesis. To do so, we have varied the parameters $\alpha_{00}$ and $\alpha_{11}$ from 0.7 to 1. The parameter $|\alpha_{10}|$ has been varied from 0 to 0.2. The phase $\phi_{10}$, associated with $\alpha_{10}$, has been varied in its total range of $[-180^\circ:180^\circ ]$. These are the only non-unitary parameters that matters in the $\nu_\mu \to \nu_e$ oscillation or $\nu_\mu \to \nu_\mu$ survival probabilities~\cite{Escrihuela:2016ube}. Therefore, all other $\alpha$ parameters have been kept constant at their boundary values.
Moreover, only those values of the parameters were chosen, for which $|\alpha_{10}|\leq \sqrt{(1-\alpha_{00}^{2})(1-\alpha_{11}^{2})}$ bound is obeyed \cite{Antusch_2014, Escrihuela:2016ube}.

\subsection{Individual analyses}
\begin{figure}[htbp]
\centering
\includegraphics[width=0.85\textwidth, trim={3.3cm 0.5cm 2.5cm 0},clip]{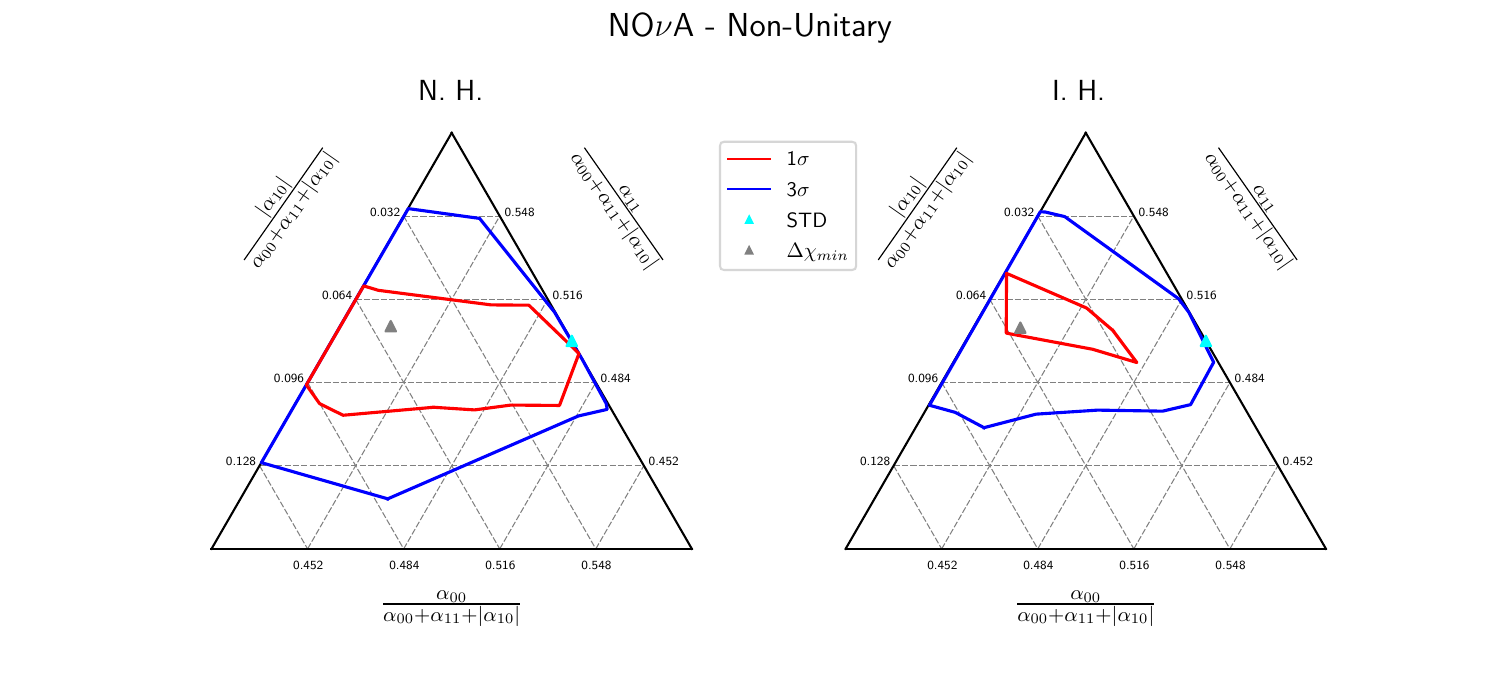}
\includegraphics[width=0.85\textwidth, trim={3.3cm 0.5cm 2.5cm 0},clip]{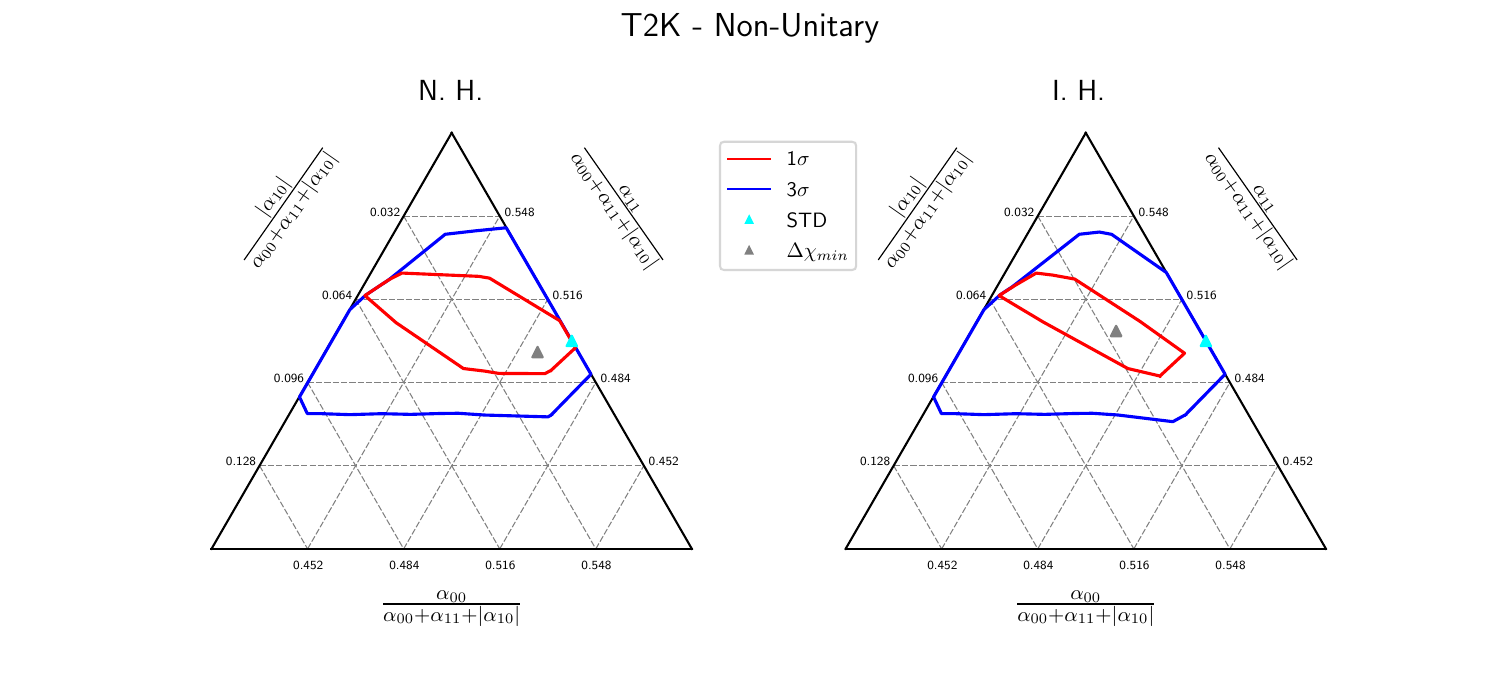}
\caption{\footnotesize{Allowed regions of the $\alpha$ parameters in a triangle plot for \nova (T2K) in the upper (lower) panel. The left (right) panel is for NH (IH). The parameter values at the best fit point are listed in Table~\ref{best-fit_tab_nova}
(\ref{best-fit_tab_t2k}) for \nova (T2K). These results are for 2019 data.}}
\label{triangle1}
\end{figure}

For the non-unitary case, the minimum $\chi^2$ for the \nova and T2K data are 44.32 and 121.37 for 50 (46 d.o.f.) and 104 (100 d.o.f.) bins respectively and they both occur at the NH. Since  We have done similar analyses for both the experiments separately with the unitary $3\times 3$ mixing hypothesis and got the minimum $\chi^2$ for the \nova and T2K data as 47.92 (42 d.o.f.) and 123.71 (96 d.o.f.), respectively, both for NH. We have noted down the values of the unitary and non-unitary parameters at the best fit point 
for \nova and T2K in Tables~\ref{best-fit_tab_nova} and \ref{best-fit_tab_t2k} respectively. It can be seen that 
the T2K data cannot rule out either of the two mixing schemes at $1\, \sigma$ C.L.. For the IH, the minimum $\dchsq$ for the \nova (T2K) data in the case of non-unitary mixing is 1.36 (2.49), and that for the unitary mixing is 2.7 (6.46).  In fig.~\ref{triangle1}, we have plotted the allowed region of the $\alpha$ parameters in triangular plots, showing the $1\,\sigma$ and $3\,\sigma$ contours. The grey triangle, labeled $\Delta\chi_{\rm min}$, corresponds to the best-fit point in the non-unitary case. The cyan triangle, labeled STD, corresponds to the best-fit point in the standard $3\times 3$ unitary case. 

\begin{figure}[htbp]
\centering
\includegraphics[width=0.6\textwidth]{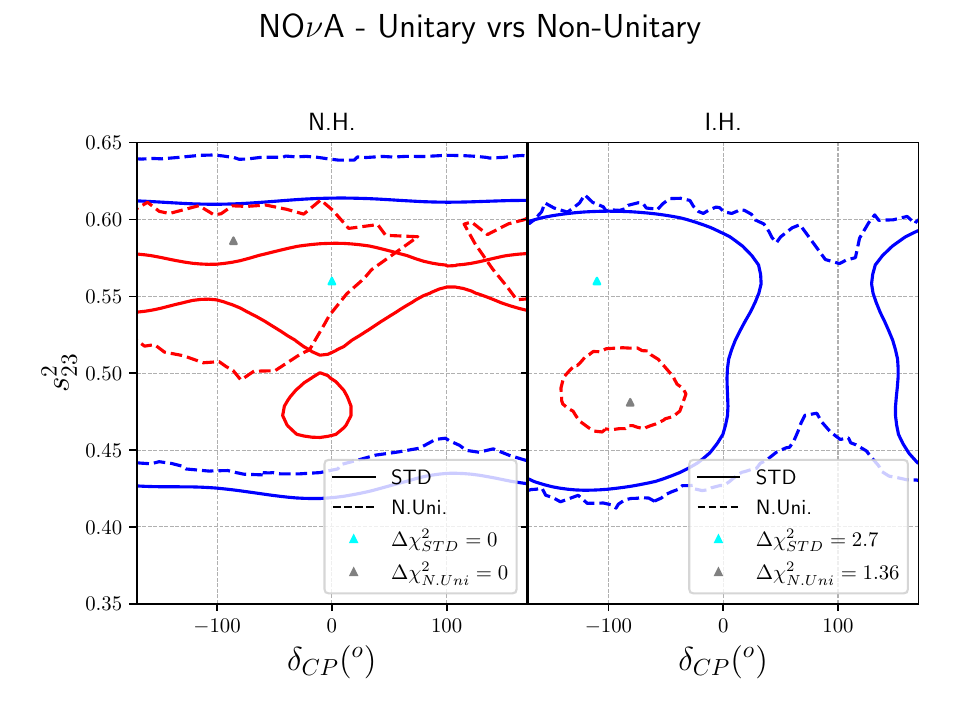}
\includegraphics[width=0.6\textwidth]{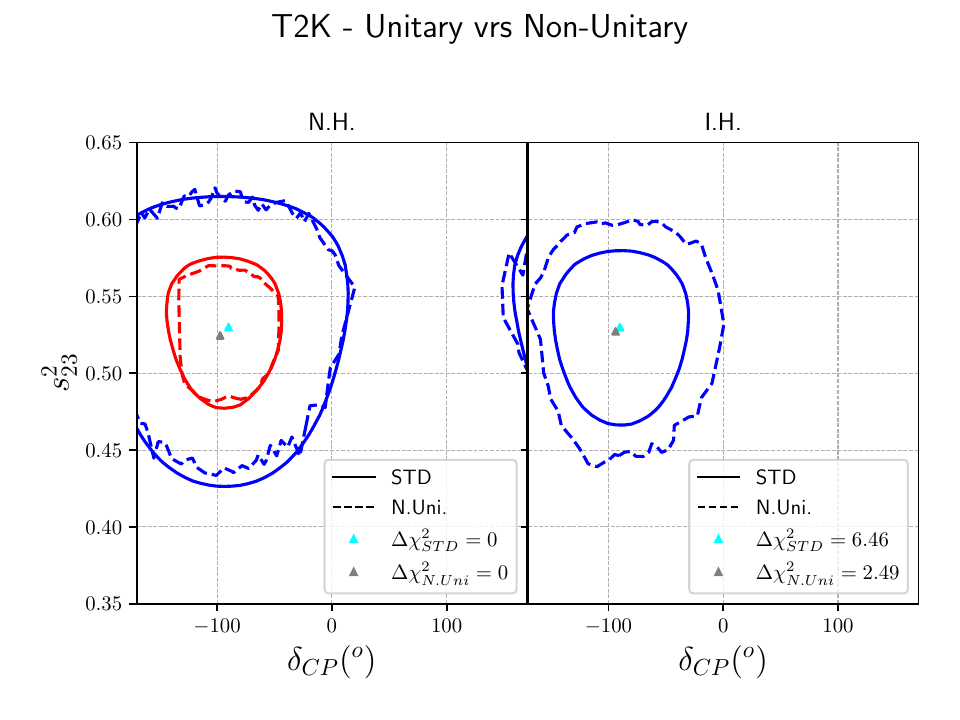}
\caption{\footnotesize{Allowed regions in the $\sin^2\tz-\dcp$ plane for \nova (T2K) in the upper (lower) panel. The left (right) panel is for NH (IH). The red (blue) lines indicate $1\, \sigma$ ($3\, \sigma$) C.L. The minimum $\chi^2$ for \nova (T2K) in the case of unitary and non-unitary mixing are 47.92 and 44.38 (123.71 and 121.37), respectively. The parameter values at the best-fit point have been mentioned in Table~\ref{best-fit_tab_nova} (\ref{best-fit_tab_t2k}) for \nova (T2K). These results are for 2019 data.}}
\label{allowed1}
\end{figure}

In fig.~\ref{allowed1}, we have shown the analysis of individual \nova (upper panels) and T2K (lower panels) data analysis in the $\dcp-\sin^2\tz$ plane for both unitary and non-unitary hypothesis.
For the standard unitary case of NO$\nu$A, we have got the exact same best-fit value as published by the collaboration \cite{Acero:2019ksn}. The allowed region is also qualitatively same as the collaboration, though we got a smaller allowed region for the CP conserving $\dcp$ values. For T2K, our best fit point is close to the collaboration best fit point \cite{Abe:2019vii} and we have got exact same allowed region as had been found by the collaboration.
 
 It is clear from the plots in fig.~\ref{allowed1} that the agreement between the two experiments in the $\sin^2\tz-\dcp$ plane is better, though not perfect, in the case of non-unitarity. In fact, the \nova data can put better constraints on the $\dcp$ values in the case of non-unitarity. While the unitary hypothesis allows the whole $\dcp$ range, the non-unitarity can exclude almost all of the UHP in the $\dcp$ range. The best-fit value of the $\sin^2 \tz$ increases a bit in the case of non-unitarity. Unlike the unitary mixing, the T2K best fit point is allowed by the \nova data, in the case of non-unitary mixing. However, the unitary hypothesis can rule out the allowed region for the IH at $1\, \sigma$ C.L., whereas in the case of non-unitary, a small region of the IH with $0.46<\sin^2 \tz<0.52$ and $-140^\circ<\dcp<-30^\circ$ is allowed with a minimum $\dchsq$ of 1.4.

In the case of T2K (fig.~\ref{allowed1}, bottom panels), the non-unitarity does not have any significant effect on the best-fit point or the allowed region, as the best-fit points are similar to the best-fit points of the unitary case. The non-unitarity just makes slightly larger significance region in the $\sin^2\tz-\dcp$ plane to be allowed at $3\, \sigma$ C.L.\ for IH. Because of that, the T2K data continue to exclude (include) the \nova best-fit point for the NH (IH) at $1\, \sigma$ ($3\, \sigma$) C.L. Unlike NO$\nu$A,  both the unitary and non-unitary mixing can exclude the IH allowed region at $1\, \sigma$ C.L.\ for the T2K data. Therefore, T2K has a better hierarchy sensitivity than \nova in the case of non-unitary hypothesis.

\begin{figure}[htbp]
\centering
\includegraphics[width=0.6\textwidth]{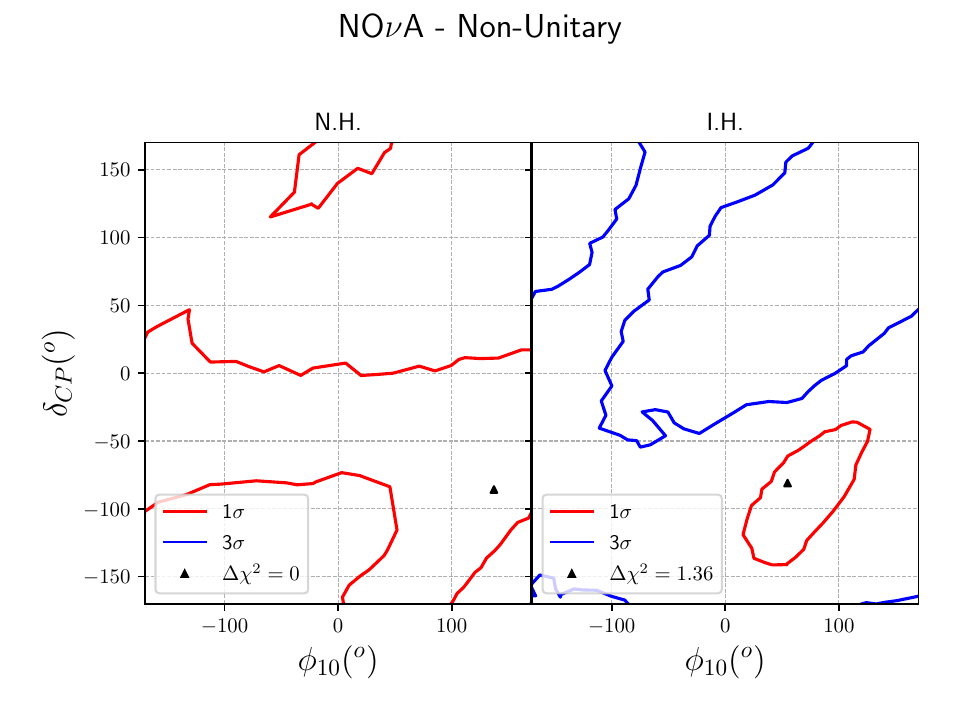}
\includegraphics[width=0.6\textwidth]{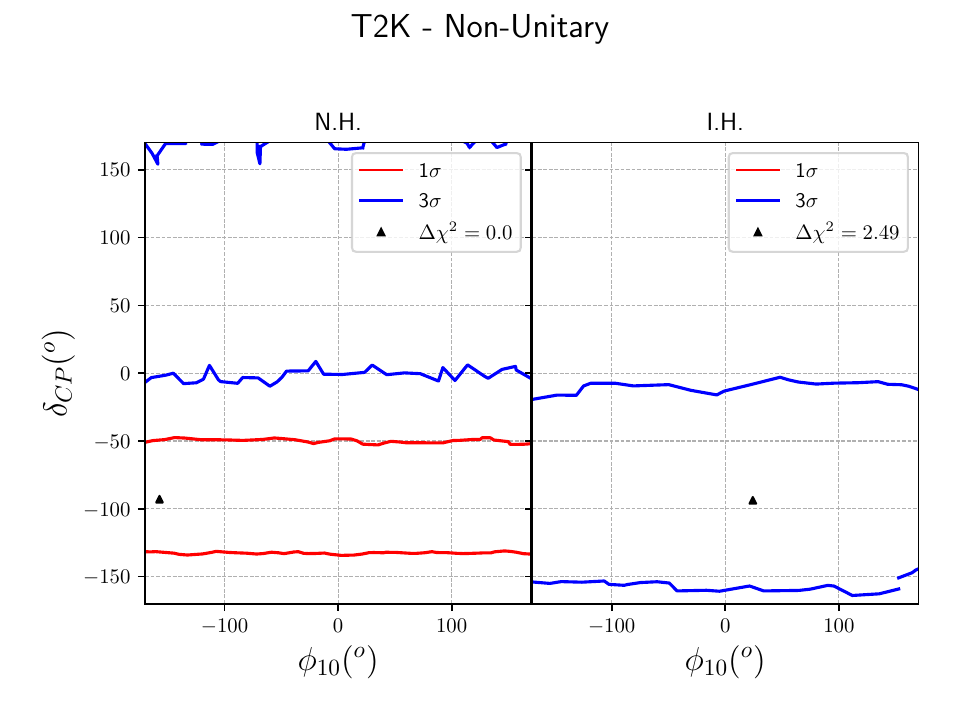}
\caption{\footnotesize{Allowed regions in the $\dcp-\phi_{10}$ plane for \nova (T2K) in the upper (lower) panel. The left (right) panel is for NH (IH). The minimum $\chi^2$ for \nova (T2K) in the case of unitary and non-unitary mixing are 47.92 and 44.38 (123.71 and 121.37), respectively. The parameter values at the best-fit point have been mentioned in Table~\ref{best-fit_tab_nova} (\ref{best-fit_tab_t2k}) for \nova (T2K). These results are for 2019 data.}}
\label{allowed2}
\end{figure}

In fig.~\ref{allowed2}, we have shown the allowed region in the $\dcp-\phi_{10}$ plane for both the \nova and T2K data. It can be seen that for the NH, both \nova and T2K cannot exclude any value of $\phi_{10}$ at $1\, \sigma$ C.L. For the IH, at $1\, \sigma$ C.L., data from \nova allow a very small region of $\dcp$ in the LHP and $\phi_{10}$ in the UHP. 

\subsection{Combined analysis}
\begin{figure}[htbp]
\centering
\includegraphics[width=0.6\textwidth]{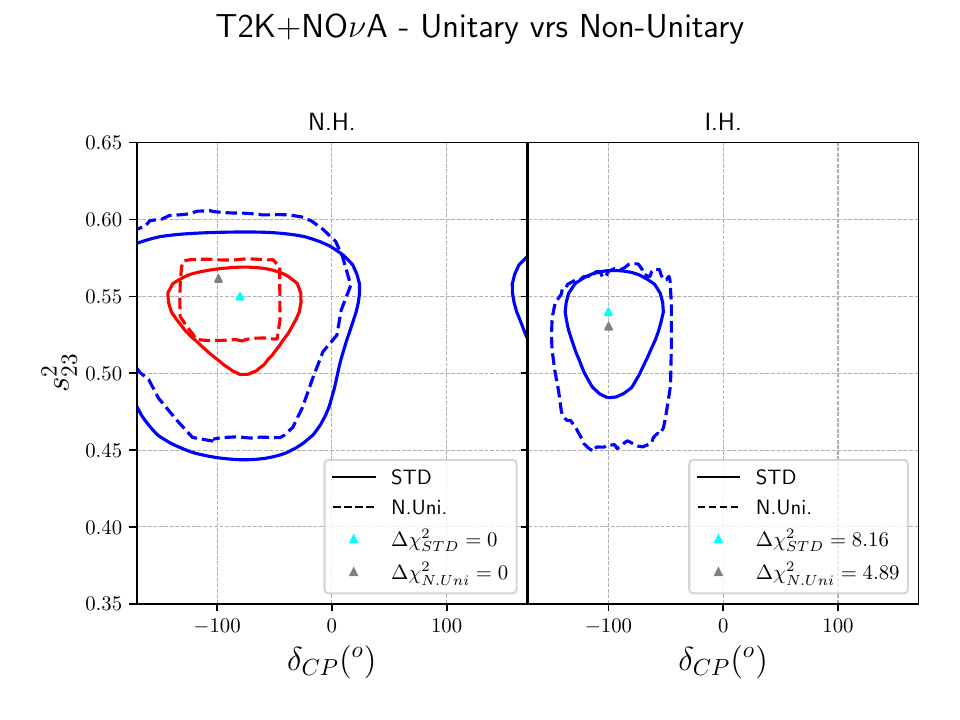}
\includegraphics[width=0.6\textwidth]{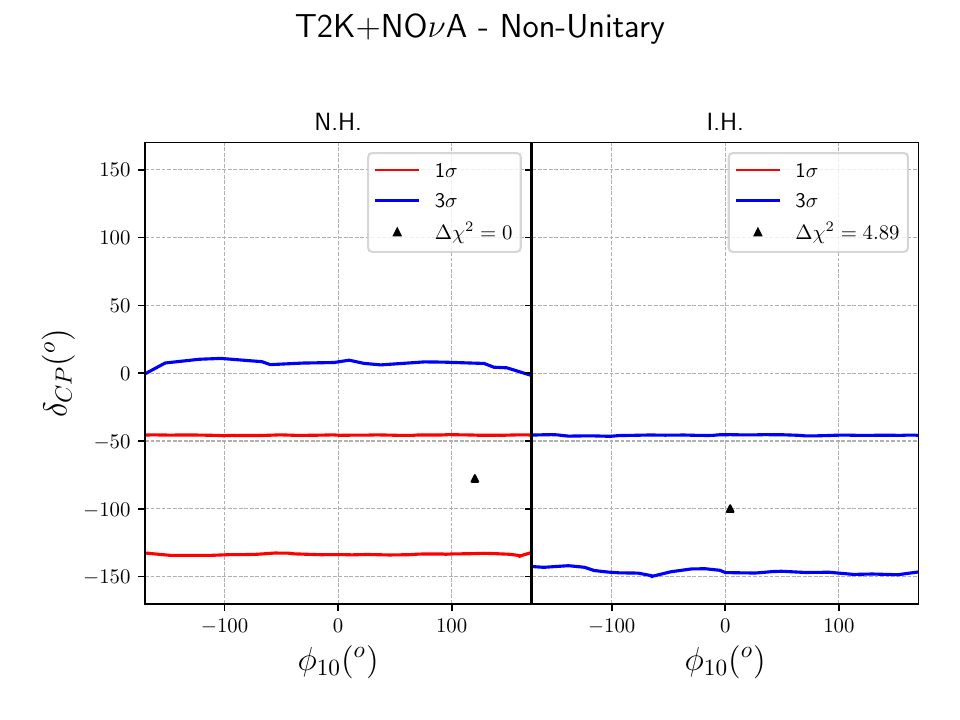}
\caption{\footnotesize{Allowed regions in the $\sin^2\tz-\dcp$ ($\dcp-\phi_{10}$) plane for the combined analysis in the upper (lower) panel. The left (right) panel is for NH (IH). The red (blue) lines indicate $1\, \sigma$ ($3\, \sigma$) C.L. The minimum $\chi^2$ for the unitary (non-unitary) case is 173.40 (170.90). The parameter values at the best-fit point have been mentioned in Table~\ref{best-fit_tab_nova+t2k}. These results are for 2019 data.}}
\label{allowed3}
\end{figure}

\begin{figure}[htbp]
\centering
\includegraphics[width=0.85\textwidth, trim={3.3cm 0.5cm 2.5cm 0},clip]{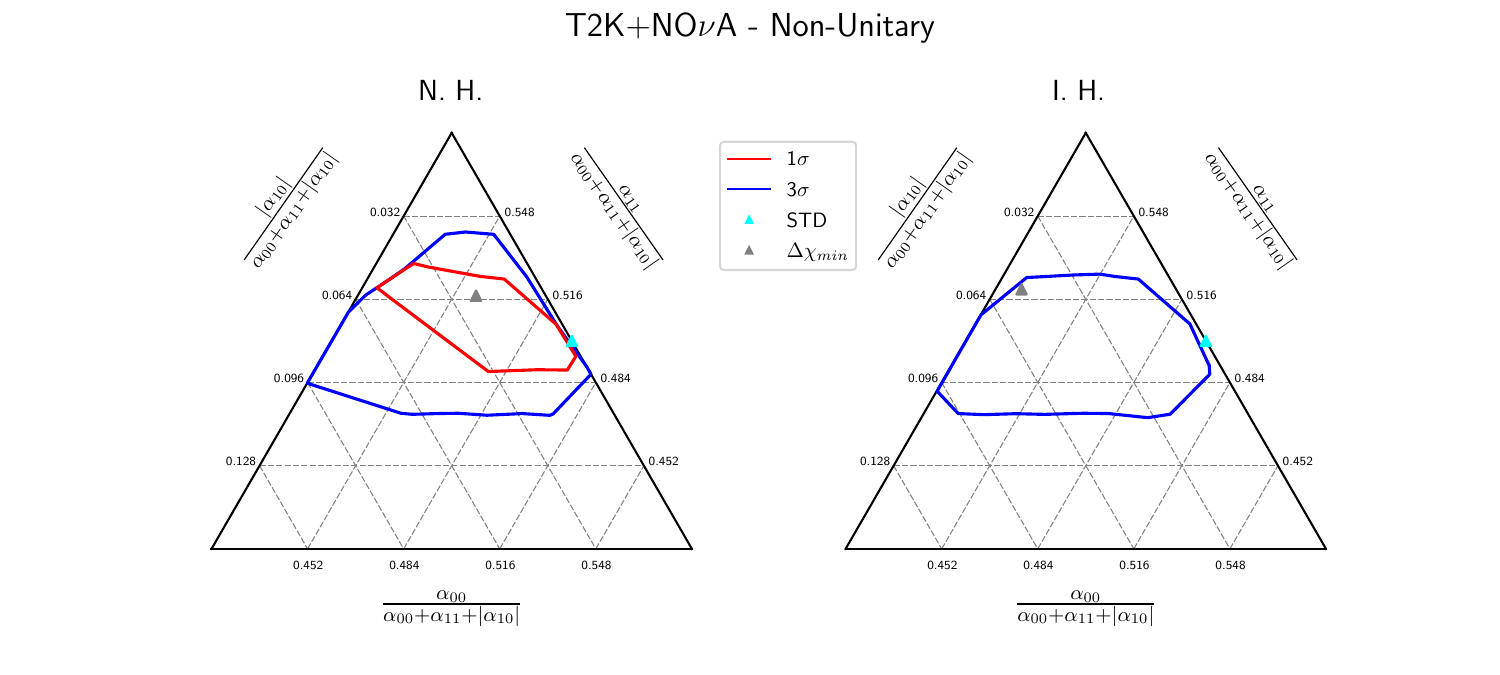}
\caption{\footnotesize{Allowed regions of the $\alpha$ parameters in a triangular plot for the combined analysis. The left (right) panel is for NH (IH). The parameter values at the best-fit point have been mentioned in Table~\ref{best-fit_tab_nova+t2k}. These results are for 2019 data.}}
\label{triangle2}
\end{figure}

Fig.~\ref{allowed3} shows the allowed region in the $\dcp-\sin^2 \tz$ and $\dcp-\phi_{10}$ planes for the combined analysis of the T2K and NO$\nu$A data. The minimum $\chi^2$ for the combined analysis with non-unitary mixing is 170.90 for 146 d.o.f. Same analysis has been done for the unitary $3\times 3$ mixing, and the minimum $\chi^2$ for the unitary case has been found out to be 173.40 for 150 d.o.f.  The combined analysis, just like NO$\nu$A, prefers non-unitary mixing at $1\, \sigma$ C.L.. The parameter values at the best-fit point for this combined analysis has been given in Table~\ref{best-fit_tab_nova+t2k}. 
The minimum $\dchsq$ for IH is 8.16 (4.89) for (non-) unitary case. Therefore in fig.~\ref{allowed3} (top right panel), there is no allowed region at $1\, \sigma$ C.L. for IH. In fig.~\ref{triangle2}, we have shown the allowed region of $\alpha$ parameters in a triangular plot, similar to fig.~\ref{triangle1}. 

\begin{figure}[htbp]
\centering
\includegraphics[width=0.65\textwidth]{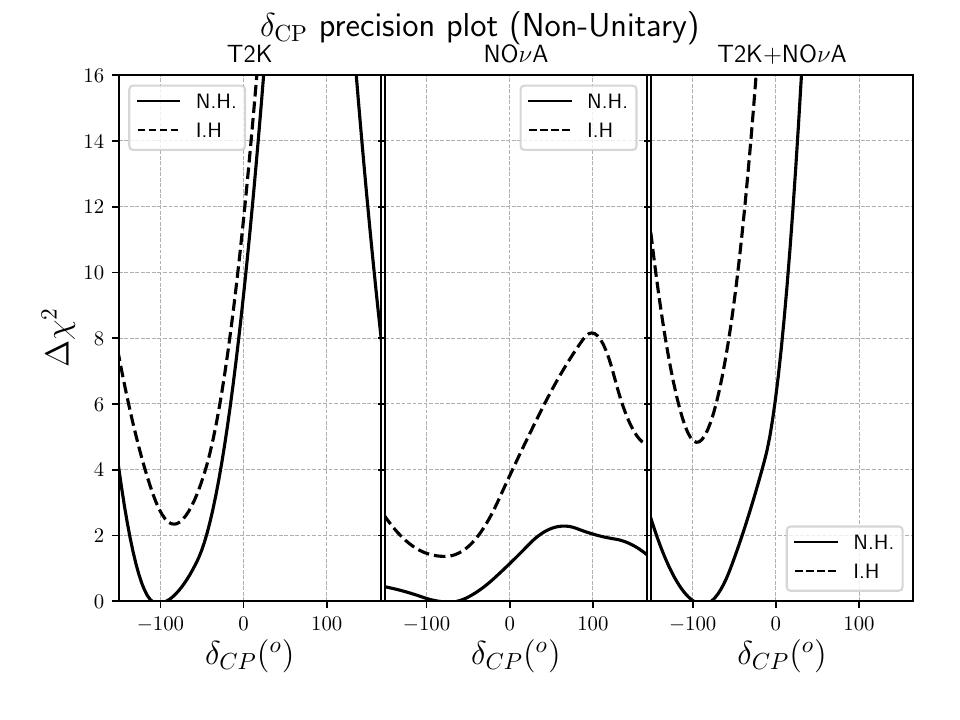}
\caption{\footnotesize{$\dchsq$ as a function of $\dcp$ for the individual T2K and NO$\nu$A, and the combined analysis. These results are for 2019 data. The limiting values of $\dchsq$ for $1\, \sigma$ and $2\, \sigma$ C.L. are $1$ and $4$ respectively.}}
\label{dcp}
\end{figure}

We have shown $\dchsq$ as a function of $\dcp$, in the case of non-unitarity, for the individual T2K and NO$\nu$A, and the combined analysis of the two experiments in fig.~\ref{dcp}. It is obvious that the individual T2K analysis can exclude $60\%$ of
the $\dcp$ plane at $2\, \sigma$ for the NH. It can also exclude the IH for $90\%$ of the $\dcp$ plane at $2\, \sigma$ C.L. But, \nova can exclude the IH only for $50\%$ of the $\dcp$ plane at $2\, \sigma$, and for the NH, it cannot disfavor any value of $\dcp$
at $2\, \sigma$ C.L. The combined analysis can exclude the IH at $2\, \sigma$ C.L. for every value of $\dcp$, and for the NH, it can exclude the UHP of $\dcp$ at $2\, \sigma$ C.L.

\begin{figure}[htbp]
\centering
\includegraphics[width=0.6\textwidth]{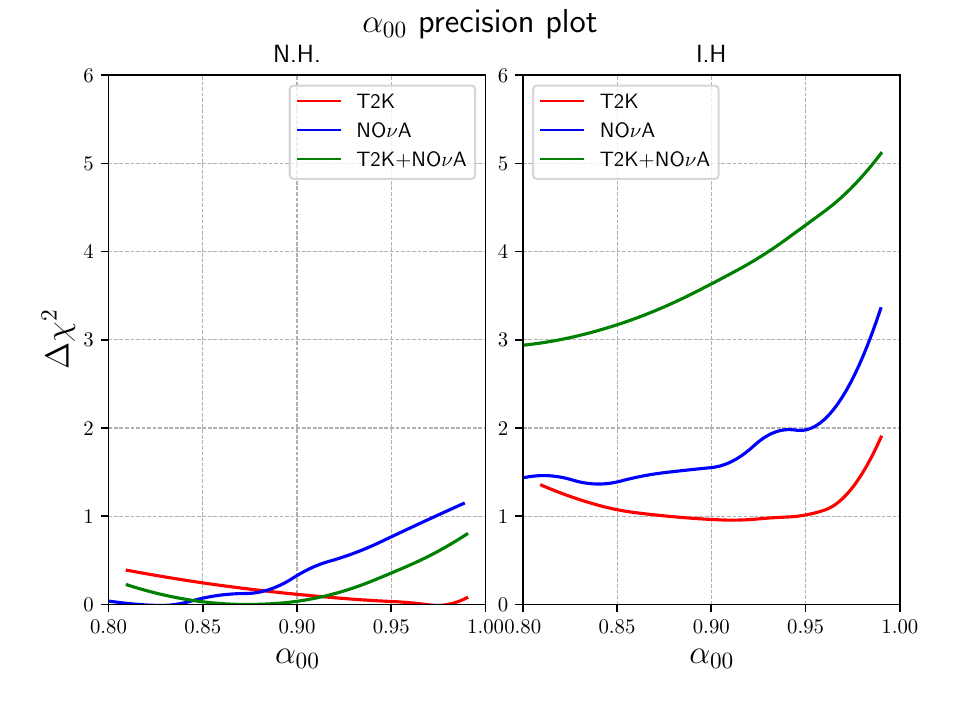}
\includegraphics[width=0.6\textwidth]{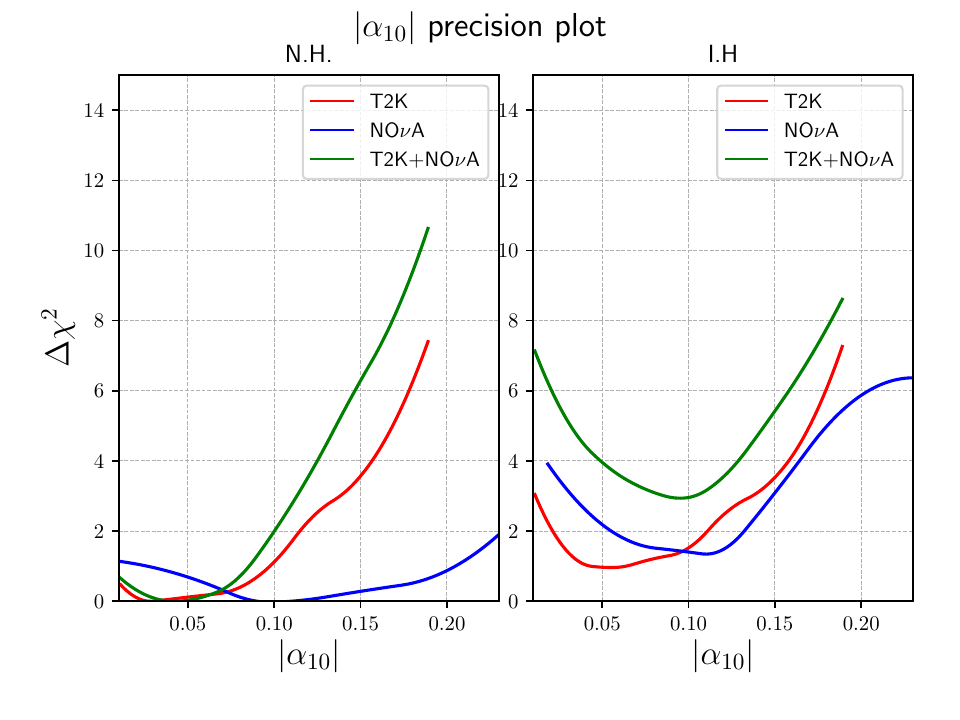}
\includegraphics[width=0.6\textwidth]{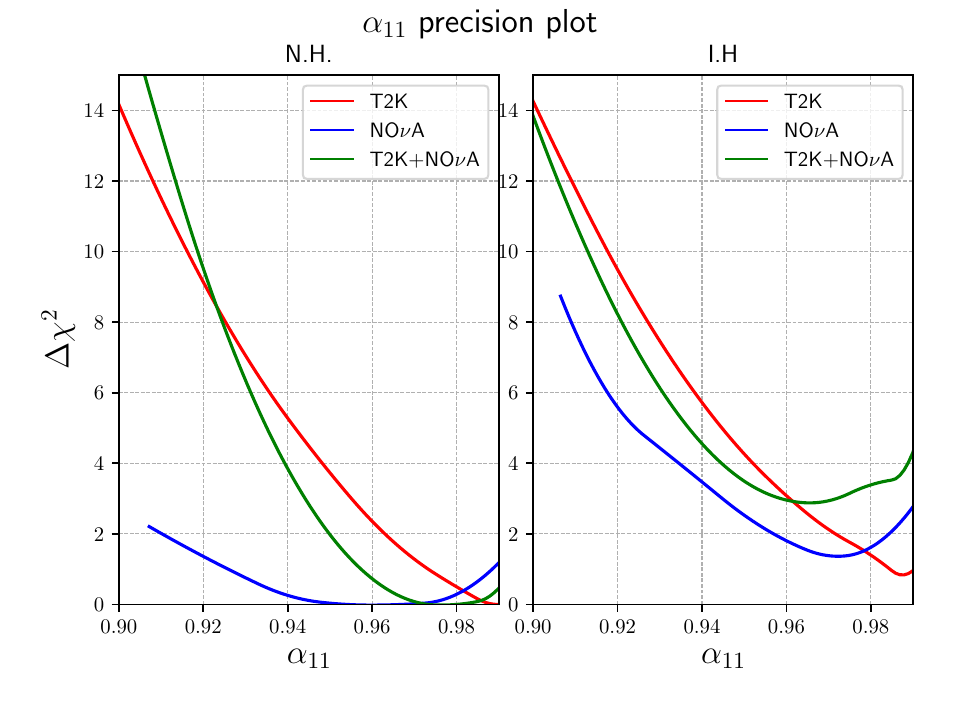}
\caption{\footnotesize{$\dchsq$ as a function of the $\alpha$ parameters for the individual T2K and NO$\nu$A, and the combined analysis. The $x$-axis of $|\alpha_{10}|$ plot is in log scale. These results are for 2019 data.}}
\label{alpha}
\end{figure}

Finally, in fig.~\ref{alpha}, we have plotted $\dchsq$ as a function of the individual non-unitary $\alpha$ parameters, so that one can have an idea of the bounds put on these parameters by the individual T2K and NO$\nu$A analyses, and their combined analysis. It is clear from  fig.~\ref{alpha} that analyses of T2K and the combined data from both the experiments prefer rather tiny deviation from unitarity at the best-fit point. However, large deviation from unitarity is allowed at $1\, \sigma$ C.L. Both \nova and the combined analysis also rules out the unitary values of $\alpha_{00}$, $|\alpha_{10}|$ and $\alpha_{11}$ at $1\, \sigma$ C.L.

We list the best-fit parameter values with $1\, \sigma$ C.L.\ intervals for \nova in Table~\ref{best-fit_tab_nova}. The best fit parameter values are reported with $1\, \sigma$ C.L.\ intervals, and the $90\%$ and $3\, \sigma$ C.L.  limit for $\alpha$ parameters from the analyses of T2K and the combined data have been mentioned in tables~\ref{best-fit_tab_t2k} and \ref{best-fit_tab_nova+t2k}, respectively.

\begin{table}

{\footnotesize
  \begin{tabular}{|l|l|l|l|l|l|l|}
    \hline
    Parameters &
      \multicolumn{2}{c|}{unitary} &
      \multicolumn{2}{c|}{non-unitary} &
       \multicolumn{2}{c|}{$90\%$} \\
    & NH & IH  & NH & IH & NH & IH \\
    \hline
   Min. $\chi^2$ (d.o.f.)  & $47.92$ (46) &  & $44.32$ (42)&  && \\
    
    &$48.65$ (46) & &$45.88$ (42)&&&\\
    
    \hline
    Min. $\dchsq$& $0$& 2.7 & 0& 1.36 && \\
    
    & $0$& 1.12 & 0& 0.66 &&\\
    
    \hline
    
    $\frac{\Delta m^{2}_{\rm eff}}{10^{-3}\, {\rm eV}^2}$  & $2.44^{+0.02}_{-0.048}$ & $-(2.44^{+0.02}_{-0.048})$ & $2.396_{-0.026}^{+0.004}$& $-(2.41_{-0.05}^{+0.01})$ && \\
    
    &$2.44^{+0.02}_{-0.048}$ & $-(2.44^{+0.02}_{-0.048})$&$2.396_{-0.026}^{+0.004}$&$-(2.41_{-0.05}^{+0.01})$&&\\
    
    \hline
     
   $\sin^2 \tz$ & $0.56_{-0.02}^{+0.01}$ & $0.56_{-0.02}^{+0.01}$ & $0.57_{-0.03}^{+0.01}$ & $0.48_{-0.02}^{+0.04}$ && \\
   &$0.59^{+0.01}_{-0.01}$&$0.59^{+0.01}_{-0.02}$&$0.62^{+0.01}_{-0.02}\oplus 0.44^{+0.01}_{-0.01}$&$0.45^{+0.01}_{-0.04}$&&
   \\
    \hline
     
   $\sin^2 2\ty$ & $0.084_{-0.002}^{+0.002}$ & $0.084_{-0.002}^{+0.003}$ &  $0.084_{-0.003}^{+0.002}$ & $0.084_{-0.003}^{+0.002}$ && \\
   &$0.084_{-0.002}^{+0.002}$&$0.084_{-0.002}^{+0.002}$&$0.084_{-0.003}^{+0.002}$&$0.084_{-0.003}^{+0.002}$&&
   \\
    \hline
       
    $\dcp/^\circ$ & $0_{-50}^{+40}$ & $-(110_{-50}^{+30})$ &  $-(72.42_{-60.55}^{+106.57})$ & $-(81.02^{+60.77}_{-30.01})$&& \\
    &$130^{+40}_{-110}$&$-(100^{+50}_{-60})$&$-(7.50^{+136.93}_{-187.50})$&$-(95.40^{+71.02}_{-43.64})$&&
    \\
    \hline
   
    $\alpha_{00}$& & & $0.83^{+0.14}_{-0.05}$ & $0.84_{-0.07}^{+0.06}$ &Out of range& Out of range \\
    &&&$0.84$&$0.72$&Out of range&Out of range\\
    \hline
    $|\alpha_{10}|$ & & & $0.107_{-0.069}^{+0.090}$ & $0.114_{-0.064}^{+0.028}$ &Out of range&$<0.18$ \\
    &&&$0.080$&$0.12^{+0.06}_{-0.12}$&Out of range& Out of range\\
    \hline
    $\alpha_{11}$ & & & $0.95^{+0.04}_{-0.03}$ & $0.97_{-0.02}^{+0.01}$ &$>0.88$&$>0.92$\\
    &&&$0.97^{+0.02}_{-0.03}$&$0.96^{+0.03}_{-0.03}$&$>0.92$&$>0.92$\\
    \hline
    $\phi_{10}/^\circ$ & & & $164.32_{-135.77}^{+15.68}$ & $54.84_{-32.57}^{+69.63}$ &&\\
    &&&$-(125.68^{+54.32}_{-305.68})$&$76.15^{+103.85}_{-86.4}$&&\\
    \hline
  \end{tabular}
  }
  \caption{Parameter values at the best-fit points for NO$\nu$A.The $1\, \sigma$ error bars have been mentioned where possible. The $90\%$ limits for 1 d.o.f have also been mentioned. In each box, the result with 2019 (2020) data has been mentioned at the top (bottom) of the box.}
  \label{best-fit_tab_nova}
\end{table}

 \begin{table}
{\footnotesize
  \begin{tabular}{|l|l|l|l|l|l|l|}
    \hline
    Parameters &
      \multicolumn{2}{c|}{unitary} &
      \multicolumn{2}{c|}{non-unitary (best-fit)} &
       \multicolumn{2}{c|}{$90\%$} \\

    & NH& IH  & NH & IH  & NH &IH \\
    \hline
    
     Min. $\chi^2$ (d.o.f.)  & $123.71$ (100) &  & $121.37$ (96)&  && \\
    
    &$95.85$ (84) & &$93.36$ (80)&&&\\
    
    \hline
    Min. $\dchsq$& $0$& 6.46 & 0& 2.49 && \\
    
    & $0$& 1.02 & 0& 0.14 &&\\
    
    \hline

    $\frac{\Delta m^{2}_{\rm eff}}{10^{-3}\, {\rm eV}^2}$  & $2.512^{+0.048}_{-0.048}$ & $-(2.512_{+0.048}^{-0.048})$ & $2.50_{-0.04}^{+0.04}$& $-(2.49_{-0.03}^{+0.05})$ & &  \\
    
     & $2.512^{+0.048}_{-0.048}$ & $-(2.512_{+0.048}^{-0.048})$ & $2.50_{-0.04}^{+0.04}$& $-(2.49_{-0.03}^{+0.05})$ & & \\
    
    \hline
     
   $\sin^2 \tz$ & $0.53_{-0.04}^{+0.03}$ & $0.53_{-0.03}^{+0.02}$ & $0.52_{-0.03}^{+0.03}$ & $0.53_{-0.03}^{+0.03}$ & &  \\
   
   &$0.55^{+0.03}_{-0.09}$&$0.56^{+0.02}_{-0.03}$&$0.46^{+0.03}_{-0.03}\oplus 0.59^{+0.03}_{-0.04}$&$0.46^{+0.03}_{-0.04}\oplus0.59^{+0.02}_{-0.02}$&&\\
   
    \hline
     
   $\sin^2 2\ty$& $0.085_{-0.002}^{+0.003}$ & $0.085_{-0.003}^{+0.002}$ &  $0.086_{-0.004}^{+0.001}$ & $0.084_{-0.001}^{+0.004}$ & &  \\
   
   & $0.085_{-0.002}^{+0.003}$ & $0.085_{-0.003}^{+0.002}$ &  $0.086_{-0.004}^{+0.001}$ & $0.084_{-0.001}^{+0.004}$ & &\\
   
    \hline
       
    $\dcp/^\circ$& $-(90_{-20}^{+30})$ & $-(90_{-20}^{+20})$ &  $-(92.88_{-30.17}^{+31.56})$ & $-(93.74^{+21.75}_{-20.46})$ & & \\
    
    &$-(100^{+50}_{-60})$&$-(90^{+30}_{-30})$&$-(71.64^{+66.36}_{-61.08})$&$-(101.26^{+45.03}_{-66.52})$&&\\
    
    \hline
   
    $\alpha_{00}$& & & $0.97$ & $0.91$ & $>0.70$ & $>0.70$ \\
    
    &&&$0.80$&$0.80$&Out of range& Out of range\\
    \hline
    $|\alpha_{10}|$ & & & $0.02$ & $0.04$ & $<0.15$ & $<0.15$ \\
    
    &&&$0.082_{-0.082}^{+0.108}$&$0.080^{+0.110}_{-0.060}$&$<0.190$&Out of range \\
    
    \hline
    $\alpha_{11}$ & & & $0.998$ & $0.997$ & $>0.95$  & $>0.96$\\
    
    &&&$0.98$&$0.98^{+0.02}_{-0.03}$&$>0.95$&$>0.95$\\
    
    \hline
    $\phi_{10}/^\circ$ & & & $-(157.27^{+22.65}_{-336.86})$ & $24.24^{+151.29}_{-178.37}$ & & \\
    
    &&&$54.77^{+97.10}_{-60.54}$&$112.69^{+42.38}_{-79.93}$&&\\
    
    \hline
  \end{tabular}
  }
  \caption{Parameter values at the best-fit points for T2K.The $1\, \sigma$ error bars have been mentioned where possible. The $90\%$ limits for 1 d.o.f have also been mentioned. In each box, the result with 2019 (2020) data has been mentioned at the top (bottom) of the box.}
  \label{best-fit_tab_t2k}
\end{table}

 \begin{table}
{\footnotesize
  \begin{tabular}{|l|l|l|l|l|l|l|}
    \hline
    Parameters &
      \multicolumn{2}{c|}{unitary} &
      \multicolumn{2}{c|}{non-unitary (best-fit)} &
       \multicolumn{2}{c|}{$90\%$} \\

    & NH& IH  & NH & IH  & NH &I H \\
    \hline
    
    Min. $\chi^2$ (d.o.f.)  & $173.40$ (150) &  & $170.90$ (146)&  && \\
    
    & & $147.14$ (134) &&$142.72$ (130)&&\\
    
    \hline
    Min. $\dchsq$& $0$& 8.16 & 0& 4.89 && \\
    
    & $1.83$& 0 & 1.07& 0 &&\\
    
    \hline

    $\frac{\Delta m^{2}_{\rm eff}}{10^{-3}\, {\rm eV}^2}$  & $2.464^{+0.024}_{-0.048}$ & $-(2.464^{+0.024}_{-0.048})$& $2.47_{-0.04}^{+0.02}$& $-(2.449_{-0.023}^{+0.003})$ & &  \\
    
    & $2.464^{+0.024}_{-0.048}$ & $-(2.464^{+0.024}_{-0.048})$& $2.47_{-0.04}^{+0.02}$& $-(2.449_{-0.023}^{+0.003})$ & &\\
    
    \hline
     
   $\sin^2 \tz$ & $0.55_{-0.02}^{+0.03}$ & $0.54_{-0.03}^{+0.01}$ & $0.55_{-0.02}^{+0.02}$ & $0.53_{-0.02}^{+0.01}$ & &  \\
   
   &$0.58^{+0.01}_{-0.02}$&$0.58^{+0.01}_{-0.02}$&$0.46^{+0.01}_{-0.02}\oplus 0.63^{+0.00}_{-0.03}$&$0.45^{+0.02}_{-0.02}$&&\\
   
    \hline
     
   $\sin^2 2\ty$& $0.085_{-0.002}^{+0.003}$ & $0.085_{-0.004}^{+0.001}$ &  $0.084_{-0.002}^{+0.003}$ & $0.085_{-0.002}^{+0.001}$ & &  \\
   
   & $0.085_{-0.002}^{+0.003}$ & $0.085_{-0.004}^{+0.001}$ &  $0.084_{-0.002}^{+0.003}$ & $0.085_{-0.002}^{+0.001}$ & &\\
   
    \hline
       
    $\dcp/^\circ$& $-(80_{-30}^{+40})$ & $-(100_{-20}^{+10})$ &  $-(77.60_{-31.05}^{+48.44})$ & $-(99.82^{+9.08}_{-23.50})$ & & \\
    
    &$-(170^{+10}_{-40})$&$-(90^{+30}_{-30})$&$-(134.56^{+44.92}_{-79.75})$&$-(102.39^{+44.68}_{-47.60})$&&\\
    
    \hline
   
    $\alpha_{00}$& & & $0.88$ & $0.80$ & $>0.72$ & $>0.70$ \\
    
    &&&$0.70$&$0.76$&Out of range& Out of range\\
    \hline
    $|\alpha_{10}|$ & & & $0.04$ & $0.09$ & $<0.12$ & $<0.12$ \\
    
    &&&$0.125^{+0.025}_{-0.085}$&$0.110^{+0.040}_{-0.070}$&$<0.17$&$<0.155$\\
    
    \hline
    $\alpha_{11}$ & & & $0.998$ & $0.97$ & $>0.94$&  $>0.95$\\
    
    &&&$0.98^{+0.01}_{-0.02}$&$0.98^{+0.01}_{-0.02}$&$>0.95$&$>0.95$\\
    
    \hline
    $\phi_{10}/^\circ$ & & & $120.41^{+59.57}_{-300.33})$ & $4.31^{+162.71}_{-181.51}$ & & \\
    
    &&&$97.24_{-74.68}^{+82.76}$&$83.27_{-40.09}^{+65.36}$&&\\
    
    \hline
  \end{tabular}
  }
  \caption{Parameter values at the best-fit points for the combined analysis of NO$\nu$A and T2K.The $1\, \sigma$ error bars have been mentioned where possible. The $90\%$ limits for 1 d.o.f have also been mentioned. In each box, the result with 2019 (2020) data has been mentioned at the top (bottom) of the box.}
  \label{best-fit_tab_nova+t2k}
\end{table}

\section{Analysis of the 2020 data}
In June, 2020, \nova \cite{Himmel:2020} and T2K \cite{Dunne:2020} have published their new data in the Neutrino 2020 conference. So far, \nova data have been analysed for $1.36 \times 10^{21}$ ($1.25 \times 10^{21}$) POT in $\nu$ ($\bar{\nu}$) mode. T2K data have been analysed for $1.97 \times 10^{21}$ ($1.63 \times 10^{21}$) POT in $\nu$ ($\bar{\nu}$) mode. According to the present data, the tension between the two experiments are even stronger. The best-fit point for \nova (T2K) is $\sin^2\tz= 0.57$ ($0.528$) and $\dcp=0.82 \pi$ ($-1.6\pi$). Moreover, there is no overlap between the $1\, \sigma$ allowed regions of the two experiments. In such a scenario, it is even more important to test new physics hypotheses with the new data from T2K and NO$\nu$A experiments. 

As before, in the GLoBES software we have tuned the signal and background efficiencies according to the Monte-Carlo simulations given by the collaborations. This time, GLoBES in its latest update has included data analysis facility and we have used GLoBES completely to analyse the data. 

To analyse the 2020 data, in eq.~(\ref{res}), for NO$\nu$A, we have used $\alpha=0.11\, (0.09)$, $\beta=\gamma=0$ for electron (muon) like events. For T2K, we have used $\alpha=0$, $\beta=0.075$, $\gamma=0.05$ for both electron and muon like events. 

For both NO$\nu$A and T2K, we have used
\begin{itemize}
    \item $5\%$ normalisation and $5\%$ energy calibration systematic uncertainties for the $e$-like events, and
    \item $5\%$ normalisation and $0.01\%$ energy calibration systematic uncertainties for the $\mu$-like events. 
    \end{itemize}

Implementing systematic uncertainties has been discussed in details in GLoBES manual \cite{Huber:2004ka, Huber:2007ji}. Here also we have used more than 30 million test data points in the parameter ranges discussed in section \ref{simulations}.

\begin{figure}[htbp]
\centering
\includegraphics[width=1.0\textwidth]{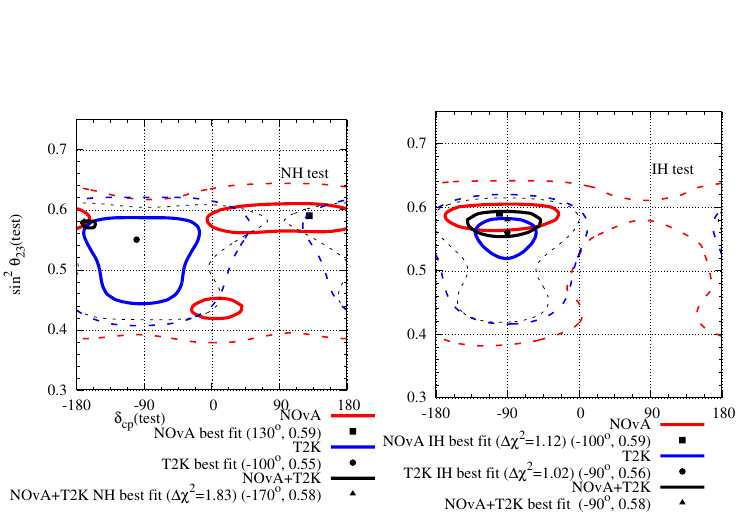}
\caption{\footnotesize{Allowed region in the $\sin^2 \tz-\dcp$ plane after analysing \nova and T2K complete 2020 data sets with unitary mixing hypothesis. The left (right) panel represents the test hierarchy as NH (IH). The red (blue) lines indicate the results for \nova (T2K) and the black lines indicate the combined analysis of both. The solid (dashed) lines indicate the $1\, \sigma$ ($3\, \sigma$) allowed regions. The minimum $\chi^2$ for \nova (T2K) with 46 (84) d.o.f. is 48.65 (95.85) and it occurs at NH. For the combined analysis, the minimum $\chi^2$ with 134 d.o.f. is 147.14 and it prefers IH. 
}}
\label{uni-2020}
\end{figure}

At first we have analysed the data with unitary mixing hypothesis and the results have been shown in fig.~\ref{uni-2020}. The minimum $\chi^2$ for \nova (T2K) with 46 (84) d.o.f. is 48.65 (95.85) and it occurs at NH. For the combined analysis, minimum $\chi^2$ for 138 bins (134 d.o.f.) is 147.14 and it occurs at IH. From fig.~\ref{uni-2020}, it can be seen that the tension between the two experiments continue. Each experiment excludes the other's allowed region at $1\, \sigma$ C.L. The combined analysis prefers IH over NH. It allows NH for a very tiny CP conserving region at $1\, \sigma$ C.L.

\begin{figure}[htbp]
\centering
\includegraphics[width=1.0\textwidth]{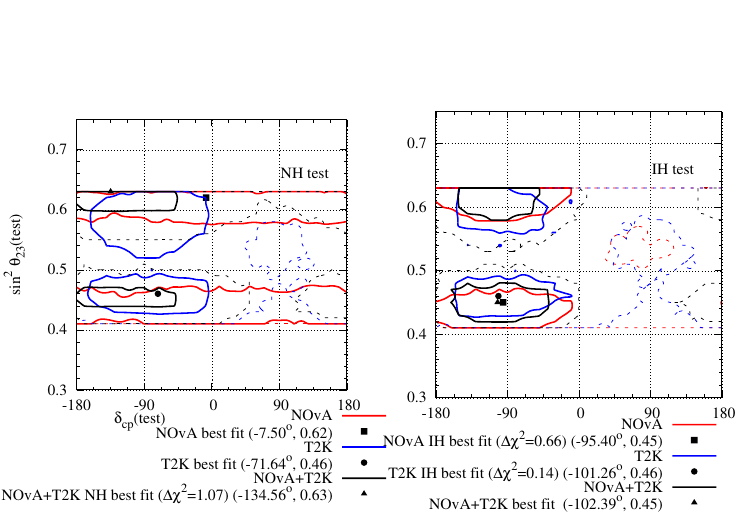}
\caption{\footnotesize{Allowed region in the $\sin^2 \tz-\dcp$ plane after analysing \nova and T2K complete 2020 data set with non-unitary mixing hypothesis. The left (right) panel represents test hierarchy NH (IH). The red (blue) lines indicate the results for \nova (T2K)
and the black line indicates the combined analysis of both. The solid (dashed) lines indicate the $1\, \sigma$ ($3\, \sigma$)
allowed regions. The minimum $\chi^2$ for \nova
(T2K) with 42 (80) d.o.f. is 45.88 (93.36) and it occurs at NH. For the combined analysis, the minimum $\chi^2$ with 130 d.o.f. is 142.72 and it prefers IH.
}}
\label{non-uni-2020}
\end{figure}

 At the next step, we have analysed the data with non-unitary hypothesis. The minimum $\chi^2$ for \nova (T2K) is 45.88 (93.36) 42 (80) d.o.f. and it is at NH. The minimum $\chi^2$ for the combined analysis, however is at IH and its value is 142.72 for 130 d.o.f. The IH best fit points for individual analysis and the NH best fit point for the combined analysis are mentioned in fig.~\ref{non-uni-2020}.

\begin{figure}[htbp]
\centering
\includegraphics[width=0.5\textwidth]{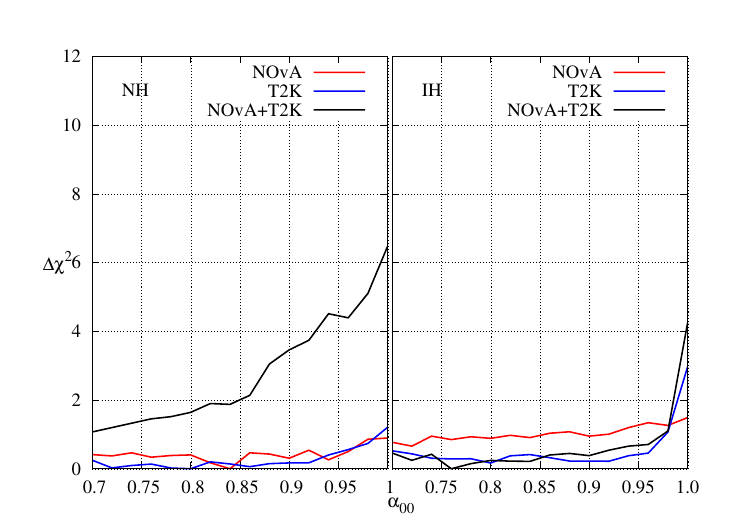}
\includegraphics[width=0.5\textwidth]{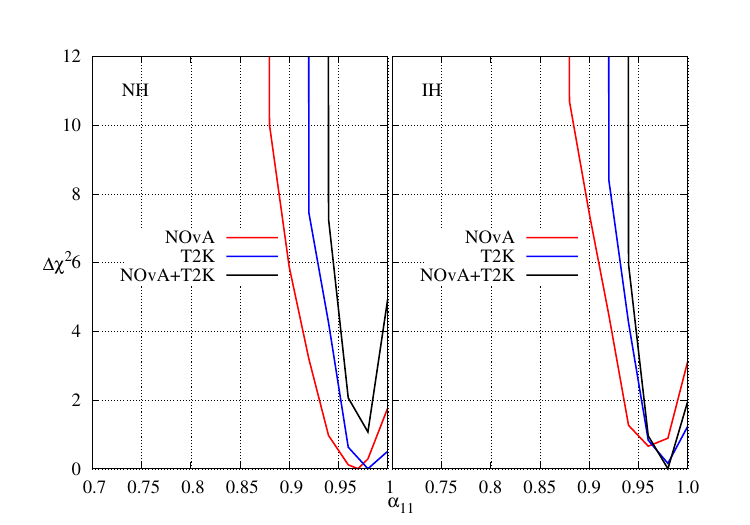}
\includegraphics[width=0.5\textwidth]{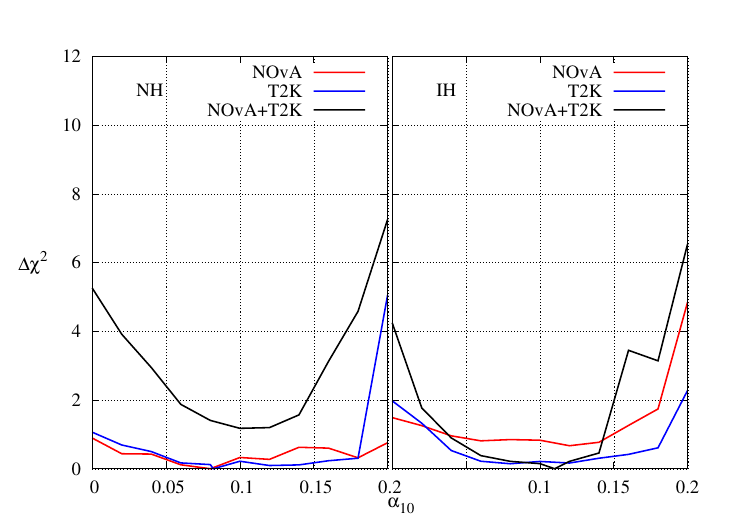}
\includegraphics[width=0.5\textwidth]{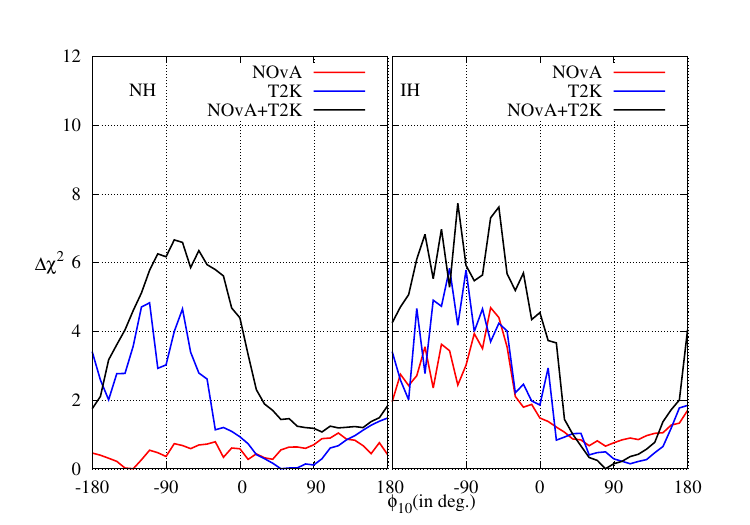}
\caption{\footnotesize{$\dchsq$ as a function of individual non-unitary parameters for 2020 long baseline data.
}}
\label{alpha-2020}
\end{figure}

In fig.~\ref{alpha-2020}, we have shown $\dchsq$ as a function of individual non-unitary parameters. To do this, we have marginalised $\dchsq$ over all parameters, except the one against which we have plotted.

It is quite certain that with the new data, both T2K and \nova prefer non-unitary mixing over unitary mixing at $1\, \sigma$ C.L.  The combined analysis excludes unitary mixing at more than $2\, \sigma$ C.L. In all three cases, the preference for non-unitarity is stronger compared to the 2019 data set. The tension between the two experiments is also reduced when analysed with non-unitary hypothesis, as there are overlaps between the allowed regions in the $\sin^2\tz-\dcp$ plane at $1\, \sigma$. However, The $\dcp$ best-fit points are still far away from each other and there is a new, mild tension between the $\tz$ octant at best-fit points between the two experiments, T2K prefers the lower octant while \nova prefers the higher octant. But \nova (T2K) cannot rule out lower (higher) octant at $1\,\sigma$ C.L. Although, NH is preferred over IH by both the experiments in separate analyses, there is an almost degenerate IH best-fit point with $\dchsq=0.66$ ($0.14$) for \nova (T2K). The combined analysis prefers IH over NH at $1\, \sigma$ C.L. It should also be noted that the two experiments have strong agreement for IH best-fit point with $\dcp$ in the LHP and $\tz$ in the lower octant. 

The best-fit points along with the $1\,\sigma$ error bars and the $90\%$ confidence level constraints on the new physics parameters have been mentioned in tables \ref{best-fit_tab_nova}, \ref{best-fit_tab_t2k} and \ref{best-fit_tab_nova+t2k}. It can be seen the constraints on the non-unitary parameters are even weaker for the 2020 data compared to the 2019 data.

\section{Conclusions}
With the 2019 data, the \nova experiment disfavors the unitary mixing at $~1\, \sigma$ C.L.\ in favor of the non-unitary mixing. The T2K experiment, however, cannot exclude any of the two hypotheses at $1\, \sigma$ C.L.  With the non-unitary hypothesis, \nova includes the T2K best-fit point at $1\,\sigma$ C.L., but T2K still continues to disfavor the \nova best-fit point at $1\, \sigma$ C.L. Unitary hypothesis can exclude the IH at $1.5\, \sigma$ C.L.\ for \nova and at $2\, \sigma$ C.L. for T2K. With the non-unitary hypothesis, hierarchy can be determined only at $1\, \sigma$ C.L.\ for the NO$\nu$A data. However, T2K can determine hierarchy at $2\, \sigma$ for $90\%$ of the $\dcp$ plane with the non-unitary hypothesis. 
The combined analysis prefers non-unitary mixing over unitary mixing at $1\, \sigma$ C.L. 

It should be noted that the tension between \nova and T2K 2019 data is reduced when both the experiments are analysed with non-unitary hypothesis. The $90\%$ confidence level limit on the $\alpha$ parameters, from these two long baseline accelerator based neutrino experiments is weaker compared to the constraints given from the global analysis in ref.~\cite{Escrihuela:2016ube} for the neutrinos only. 
 
 For the latest 2020 data set, both the experiments individually prefer non-unitary mixing over the unitary one at $1\, \sigma$ C.L., preferring NH in both cases. The combined analysis, however, prefers IH both for unitary and non-unitary mixing but the non-unitary mixing is favored over unitary mixing at more than $2\, \sigma$ C.L. The tension between the two experiments is also reduced when analysed with non-unitary hypothesis. Both the experiments lose $\tz$ octant sensitivity when analysed with non-unitarity.
 The constraints on the non-unitary parameters are even weaker with the 2020 data as compared to the 2019 data. As a consequence, the preference for non-unitarity is stronger with the 2020 data.

It can be commented that the present long baseline data prefer non-unitary mixing giving a hint of the possibility of new physics. It is important that the future analysis of \nova and T2K data are done with the non-unitary $3\times 3$ mixing hypothesis, besides the standard unitary mixing hypothesis in order to find new physics signatures. 
If the two experiments continue to have better agreement with the non-unitary hypothesis, that can be a strong hint of the presence of new physics in the neutrino sector. 

\section*{Acknowledgement}
 PP thanks to the CNPq funding grant 155374/2018-4, FAPESP funding grant 2014/19164-6 and the partial support of the Coordena\c{c}\~ao de Aperfei\c{c}oamento de Pessoal de N\'ivel Superior - Brasil (CAPES) - Finance Code 001.
 SR acknowledges support from the National Research Foundation (South Africa) with grant No. 111749 (CPRR) and by the University of Johannesburg Research Council grant. SR also thanks Max Planck Institut fu\"r Kernphysik, Heidelberg, Germany for hospitality and A.Yu. Smirnov for useful discussion where part of this work was done.


\bibliographystyle{apsrev}

 \bibliography{referenceslist}

\end{document}